\documentclass[%
 reprint,
 superscriptaddress,
 amsmath,amssymb,amsthm,
 aps,
floatfix,
]{revtex4-2}

\newcommand{\topic}[1]{\textbf{#1}}

\newcommand{\aref}[1]{Appendix~\ref{#1}}
\newcommand{\eref}[1]{Eq.~(\ref{#1})}
\newcommand{\tref}[1]{Table.~\ref{#1}}
\newcommand{\fref}[1]{Fig.~\ref{#1}}

\newcommand{\panel}[1]{(#1)}
\newcommand{\cpanel}[1]{\textbf{(#1)}}

\usepackage{graphicx}% Include figure files
\usepackage{dcolumn}% Align table columns on decimal point
\usepackage{bm}% bold math
\usepackage[colorlinks]{hyperref}% add hypertext capabilities
\hypersetup{%
    plainpages=true,
    breaklinks=true,
    hypertexnames=false,
    pageanchor=true,
    colorlinks=true,
    linkcolor={blue},
    citecolor={blue},
    urlcolor={blue},
    anchorcolor={black}
}
\usepackage{hhline}

\usepackage{braket, siunitx}

\begin{document}

\preprint{APS/123-QED}

% \title{Circularly Polarized Driving and Commensurate Pulses \\ for Fast Single-Qubit Gates with Fluxonium}
\title{Suppressing Counter-Rotating Errors for Fast Single-Qubit Gates with Fluxonium}

\def\RLEaffil{Research Laboratory of Electronics, Massachusetts Institute of Technology, Cambridge, MA 02139, USA}
\def\LLaffil{Lincoln Laboratory, Massachusetts Institute of Technology, Lexington, MA 02421, USA}
\def\Physaffil{Department of Physics, Massachusetts Institute of Technology, Cambridge, MA 02139, USA}
\def\EECSaffil{Department of Electrical Engineering and Computer Science, Massachusetts Institute of Technology, Cambridge, MA 02139, USA}
\def\affilAQ{\textit{Atlantic Quantum, Cambridge, MA 02139}}
\def\affilHRL{\textit{HRL Laboratories, Malibu, CA 90265}}
\def\affilGoogle{\textit{Google Quantum AI, Goleta, CA 93111}}

\author{David A. Rower}
\email[These authors contributed equally; ]{rower@mit.edu}
\affiliation{\Physaffil}
\affiliation{\RLEaffil}

\author{Leon Ding}
% \email{leonding@mit.edu}
\altaffiliation[These authors contributed equally; Present address: ]{\affilAQ}
\affiliation{\Physaffil}
\affiliation{\RLEaffil}

\author{Helin~Zhang}
\affiliation{\RLEaffil}

\author{Max~Hays}
\affiliation{\RLEaffil}

\author{Junyoung~An}
\affiliation{\RLEaffil}
\affiliation{\EECSaffil}

\author{Patrick~M.~Harrington}
\altaffiliation[Present address: ]{\affilHRL}
\affiliation{\RLEaffil}

\author{Ilan~T.~Rosen}
\affiliation{\RLEaffil}

\author{Jeffrey~M.~Gertler} 
\affiliation{\LLaffil} 

\author{Thomas~M.~Hazard} 
\affiliation{\LLaffil} 

\author{Bethany~M.~Niedzielski} 
\affiliation{\LLaffil} 

\author{Mollie~E.~Schwartz}
\affiliation{\LLaffil}

\author{Simon~Gustavsson} 
\altaffiliation[Present address: ]{\affilAQ}
\affiliation{\RLEaffil}

\author{Kyle~Serniak} 
\affiliation{\RLEaffil}
\affiliation{\LLaffil} 

\author{Jeffrey~A.~Grover}
\affiliation{\RLEaffil}

\author{William~D.~Oliver}
\email{william.oliver@mit.edu}
\affiliation{\Physaffil} 
\affiliation{\RLEaffil} 
\affiliation{\EECSaffil} 
% \affiliation{\LLaffil} 

\date{\today}

\begin{abstract}
Qubit decoherence unavoidably degrades the fidelity of quantum logic gates.
Accordingly, realizing gates that are as fast as possible is a guiding principle for qubit control, necessitating protocols for mitigating error channels that become significant as gate time is decreased.
One such error channel arises from the counter-rotating component of strong, linearly polarized drives.
This error channel is particularly important when gate times approach the qubit Larmor period and represents the dominant source of infidelity for sufficiently fast single-qubit gates with low-frequency qubits such as fluxonium.
In this work, we develop and demonstrate two complementary protocols for mitigating this error channel.
The first protocol realizes circularly polarized driving in circuit quantum electrodynamics (QED) through simultaneous charge and flux control.
% We demonstrate the generation of tunable-polarization drives and implement a co-rotating drive for use in gates.
The second protocol---commensurate pulses---leverages the coherent and periodic nature of counter-rotating fields to regularize their contributions to gates, enabling single-qubit gate fidelities reliably exceeding $99.997\%$.
This protocol is platform independent and requires no additional calibration overhead.
%We reliably achieve single-qubit gate fidelities in excess of $99.997\%$.
This work establishes straightforward strategies for mitigating counter-rotating effects from strong drives in circuit QED and other platforms, which we expect to be helpful in the effort to realize high-fidelity control for fault-tolerant quantum computing.
\end{abstract}

\maketitle

\section{Introduction}

% Superconducting qubits have emerged as a leading contender for performing quantum computation, with steadily increasing circuit sizes and rising gate fidelities approaching the levels required to implement and scale error correcting codes~\cite{Nakamura1999, Koch2007, Barends2014, Arute2019, Acharya2023}.
Superconducting qubits have emerged as a leading contender for performing quantum computation~\cite{divincenzoPhysicalImplementationQuantum2000}, with steadily increasing circuit sizes and rising gate fidelities approaching the levels required to begin scaling error correcting codes~\cite{Nakamura1999, Koch2007, Barends2014, Arute2019, Acharya2023}.
% However, further advances are required to realize useful quantum computations. 
However, gate fidelities must improve further to realize useful quantum computations, e.g., by surpassing code thresholds to yield algorithmically-relevant logical error rates~\cite{Fowler2012, Acharya2023, Bluvstien2023, dasilva2024}.

% One central requirement for a useful quantum processor is the implementation of high-fidelity gates~\cite{divincenzoPhysicalImplementationQuantum2000}.
% The quality of gates is heavily influenced by unavoidable error due to decoherence, which is
Decoherence poses a significant challenge to the realization of high-fidelity gates, contributing an error proportional to $t_g/T_\text{coh}$, where $t_g$ represents the duration of a gate operation and $T_\text{coh}$ the relevant coherence timescale. For superconducting qubits, decoherence often limits the fidelity of state-of-the-art gates. Consequently, to build useful quantum hardware, we aim to increase system coherence and decrease gate times as much as possible. 

The speed of quantum logic gates is generally limited by effects that become significant as gate times approach relevant system timescales.
As a primary example, leakage to non-computational states becomes significant when gate times approach the timescale set by the qubit anharmonicity~\cite{Motzoi2009, Zhu2021}. %(owing to the spectral content of gate pulses overlapping with those transitions).
This example is especially relevant for transmon qubits~\cite{Koch2007, Schreier2008}, for which low anharmonicities of typically $|\alpha|/2\pi\sim\SI{200}{MHz}$ limit gate times to $(|\alpha|/2\pi)^{-1} \sim \SI{5}{ns}$~\cite{chenMeasuringSuppressingQuantum2016, hyyppaReducingLeakageSinglequbit2024}.
As another example, linear drives~\cite{blochMagneticResonanceNonrotating1940} for gates with durations approaching the qubit Larmor period ($t_g / \tau_L \rightarrow 1$) give rise to significant undesirable counter-rotating dynamics, resulting from the breakdown of the rotating-wave approximation (RWA)~\cite{Fuchs2009, Avinadav2014, Deng2015, Burgar2022}.
These effects are typically negligible for transmons due to their high transition frequency ($t_g / \tau_L \gtrsim \SI{5}{ns} \cdot \SI{4}{GHz} = 20$). 
% However, counter-rotating effects can pose a major challenge for fast resonant control of low-frequency qubits such as fluxonium. 
 
Both the intrinsic control limitations and coherence properties of a particular qubit are dictated by its Hamiltonian.
% To this end, the ability to engineer Hamiltonians in the context of circuit quantum electrodynamics (QED) has enabled devices comprising artificial atoms with spectrally isolated transitions that can be used to encode qubits, designed with intrinsic protection from dominant noise sources.
% Early generations of superconducting qubits, including charge~\cite{Nakamura1999, Vion2002} and persistent-current flux~\cite{Mooij1999, Orlando1999, Friedman2000, vanderWal2000} qubits, were characterized by high anharmonocities but suffered from low coherence due to offset-charge- and flux noise.
% Most recently, the transmon~\cite{Koch2007, Schreier2008} has achieved coherence on the scale of 10-100 \SI{}{\micro s} due to its reduced sensitivity to offset-charge fluctuations; however, the limitations imposed by its low anharmonicity and unprotected computational states have motivated the investigation of circuits with more favorable coherence and control properties.
The interplay between qubit anharmonicity, sensitivity to noise, and circuit simplicity for superconducting qubits has been elucidated over the past several decades~\cite{Nakamura1999, Vion2002, Mooij1999, Orlando1999, Friedman2000, vanderWal2000, Manucharyan2009, Yan2016, Yurtalan2021}, leading to the transmon as the current workhorse superconducting qubit.
However, the limitations imposed by its low anharmonicity and unprotected computational states~\cite{gyenisMovingTransmonNoiseProtected2021} have motivated further investigation of circuits with more favorable coherence and control properties.
% The transmon has achieved coherence on the scale of up to hundreds of microseconds due to its reduced sensitivity to offset-charge fluctuations; however, the limitations imposed by its low anharmonicity and unprotected computational states~\cite{gyenisMovingTransmonNoiseProtected2021} have motivated the further investigation of circuits with more favorable coherence and control properties.
The fluxonium~\cite{Manucharyan2009} is one such qubit, featuring a transition frequency typically less than \SI{1}{GHz}, with state-of-the-art coherence times and single- and two-qubit gate fidelities \cite{Pop2014, Nguyen2019, Somoroff2023, Ding2023, Zhang2024}.
One beneficial property of the fluxonium is its high anharmonicity, typically several gigahertz, when operated at a flux bias of a half magnetic flux quantum $\Phi_0 / 2$.
This property mitigates leakage during fast single-qubit gates, enabling the exploration of new dominant error channels and strategies to mitigate them.
In particular, fast, resonant control of fluxonium naturally places one in the regime $t_g / \tau_L \lesssim \SI{5}{ns} \cdot \SI{1}{GHz} = 5$, where errors due to counter-rotating effects start to become severe.
 
In this work, we explore fast single-qubit gates based on the resonant control of a fluxonium qubit in the regime of few-cycle pulses ($1 \lesssim t_g / \tau_L \lesssim 5$), where errors from counter-rotating dynamics are significant.
We develop two complementary strategies for mitigating these errors. 
Our first strategy takes inspiration from circularly polarized free-space electromagnetic fields and realizes tunable-polarization drives~\cite{London2014} with simultaneous charge and flux control.
% This approach leverages the linear flux coupling of the fluxonium Hamiltonian~\cite{Bryon2023}.
We demonstrate the tunability of drive polarization and the calibration of co-rotating drives for gates.
Our second strategy, enabling fidelities exceeding $99.997\%$ with only a linear drive, involves leveraging the time-periodicity of the counter-rotating fields to regularize their contributions to all pulses~\cite{Laucht2016}.
This approach, hereafter referred to as commensurate pulses, eliminates the coherent error channel posed by counter-rotating effects with no additional calibration overhead and can be implemented straightforwardly to mitigate these effects in any platform where fast resonant control is desired.
We utilized both approaches to implement single-qubit gates, exploring different drive schemes---charge, flux, and circularly polarized drives---and the resulting fidelity dependence on the gate duration.
Finally, we developed an error budget for our gates and investigated their stability, finding that our best gates were coherence limited, and the performance remained stable (error per gate fluctuations $\lesssim 1.13 \times 10^{-5}$) for the entire measurement duration of 34 hours after an initial calibration.

\section{Device and Theory}
In this section, we detail the device, our two complementary protocols for mitigating counter-rotating errors for fast gates, and our single-qubit gate implementation.

\subsection{Fluxonium Device}
\begin{figure}[!tb]
\includegraphics{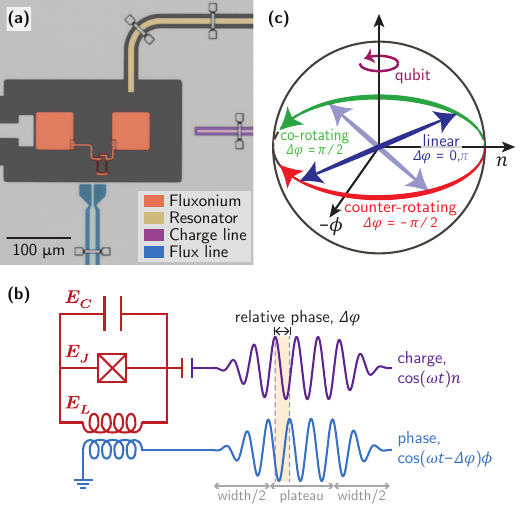}
\caption{
\label{fig:cartoon}
\textbf{Device and concept.}
\cpanel{a} False-colored optical micrograph of the fluxonium qubit (orange) with a coupled charge (purple) and flux line (blue).
\cpanel{b} Diagram illustrating how all drives are performed in this experiment.
All drives comprise a phase-sensitive linear combination of a charge and flux drive, parameterized by a cosine-shaped rise-fall and a flat top.
\cpanel{b} Bloch sphere representation of a qubit with the trajectory of various drive polarization vectors drawn.
A co-rotating (green) drive has a polarization vector rotating with the same orientation as the qubit, a counter-rotating (red) drive has a polarization vector rotating in the opposite direction as the qubit, and a linear (blue) drive has a polarization vector which traces out a line.
}
\end{figure}

Our device comprises a two-dimensional differential fluxonium qubit capacitively coupled to a charge line and inductively coupled to a flux line as shown in \fref{fig:cartoon}\panel{a}.
Modeling just the fluxonium qubit and the two drive lines, our device obeys the system Hamiltonian 
\begin{equation} \label{eq:fullHamiltonian} % Combined Hamiltonian
    \hat H = \hat{H}_0 + \hbar \Omega_c \cos(\omega_d t) \hat{n} + \hbar\Omega_f \cos(\omega_d t - \Delta \varphi) \hat{\phi},
\end{equation}
where $\hat{H}_0$ is the bare fluxonium Hamiltonian 
\begin{equation} % Base Fluxonium
    \hat H_0 = 4 E_C \hat{n}^2 + \frac{1}{2} E_L \hat{\phi}^2 - E_J \cos(\hat\phi - \phi_\text{dc}).
\end{equation}
This Hamiltonian is written in the lab frame, where $\Omega_c$ ($\Omega_f$) describes the amplitude of a cosinusoidal charge (flux) drive with frequency $\omega_d$ and phase difference $\Delta \varphi$ between the two drives. In the fluxonium Hamiltonian, $\hat{n}$ and $\hat{\phi}$ represent the charge and phase operators, $E_C/h = \SI{1.30}{GHz}$, $E_L/h=\SI{0.59}{GHz}$, and $E_J/h=\SI{5.71}{GHz}$ are the charging-, inductive-, and Josephson energy, respectively, and $\phi_\text{dc}$ is a phase offset resulting from a static external magnetic flux $\Phi_\text{dc} / \Phi_0 = \phi_\text{dc} / 2\pi$ supplied by a superconducting coil inductively coupled to the fluxonium loop.
The sample qubit was a subsystem of a device comprising two fluxonium qubits with a capacitively-coupled transmon coupler (refer to device A, fluxonium 2 of Ref.~\cite{Ding2023}).
We note that the linear flux drive arises from the allocation of time-dependent flux inside the inductor term, which has recently been theoretically and experimentally verified~\cite{You2019, Bryon2023}. 

Our experiments were performed at $\Phi_\text{dc} = 0.5\Phi_0$, where the qubit had a frequency $\omega_{01} / 2\pi \approx \SI{243}{MHz}$ (Larmor period $\tau_L \approx \SI{4.1}{ns}$) and coherence times generally between $\SI{200}{\micro s} \lesssim T_1, T_{2E} \lesssim \SI{500}{\micro s}$. 
Dispersive readout was performed with a capacitively-coupled resonator at frequency $\omega_r / 2\pi = \SI{7.08}{GHz}$, with linewidth $\kappa/2\pi \approx \SI{1.5}{MHz}$ and dispersive shift $\chi/2\pi \approx \SI{1}{MHz}$~\cite{Krantz2019}.
We herald the desired initial state with a preceding projective measurement, with a buffer time of \SI{2}{\micro s} before applying qubit pulses to avoid photon-shot-noise dephasing.
All charge and flux control pulses were directly synthesized on a high-bandwidth arbitrary waveform generator (see \aref{appendix:wiring} for details).

We emphasize the relevance of fluxonium qubits for our experiments exploring counter-rotating dynamics, as their typically low qubit frequency and high anharmonicity cause counter-rotating effects to manifest before leakage into non-computational states is observed when performing Rabi-based single-qubit gates. 

\subsection{Circularly Polarized Driving}

Here, we detail the ability to tune qubit drive polarization by controlling the relative phase $\Delta \varphi$ of simultaneous charge and flux drives [\fref{fig:cartoon}\panel{b}]. 
Truncating to the ground and first excited state manifold of the system Hamiltonian \eref{eq:fullHamiltonian} and considering charge and flux drives of the same strength (${\Omega_c \lvert\bra{0}\hat{n}\ket{1}\rvert = \Omega_f \lvert\bra{0}\hat{\phi}\ket{1}\rvert \equiv \Omega / 2}$), the two-level Hamiltonian, with $\hat{H}\ket{0} = 0$, takes the form
\begin{align*}
    \frac{\hat{H}}{\hbar} = \;\, &\omega_{01} \ket{1}\bra{1} \\
    &- i\frac{\Omega}{4} \cos(\omega_d t)(\ket{0}\bra{1} - \ket{1}\bra{0}) \\
    &+ \frac{\Omega}{4}\cos(\omega_d t - \Delta \varphi)(\ket{0}\bra{1} + \ket{1}\bra{0}).
\end{align*}
From the above equation, we see that the two drive terms act along orthogonal axes of the Bloch sphere in the lab frame.
This is a consequence of the canonically conjugate nature of the charge and flux operators, and it enables the effective field polarization to be tuned by the relative phase of the charge and flux drives.
This may be contrasted with the polarization of free-space electromagnetic fields, where polarization is defined by the relation of real-space field components.

\topic{Linear drives.}
% Setting the relative phase to $\Delta \varphi = 0$ ($\pi$) [\fref{fig:cartoon}\panel{b}, blue] yields a drive oscillating along the direction $\hat{n} + \hat{\phi}$ ($\hat{n} - \hat{\phi}$), i.e. a linearly polarized field.
Setting the relative phase to $\Delta \varphi = 0$ ($\pi$) [\fref{fig:cartoon}\panel{c}, blue] yields a linearly polarized drive with drive operator $\hat{n} + \hat{\phi}$ ($\hat{n} - \hat{\phi}$).
One can also trivially generate linear drives by applying an individual charge or flux drive (oscillating only along the $n$ or $\phi$ axes respectively).

\topic{Co-rotating drives.}
With $\Delta \varphi = \pi/2$ [\fref{fig:cartoon}\panel{c} green], the net drive field instead travels along the equator of the Bloch sphere in the same direction as the qubit Larmor precession [\fref{fig:cartoon}\panel{c} purple].
Performing a rotating-frame transformation~\cite{rabiUseRotatingCoordinates1954} with frequency $\omega_{d}$ co-rotating with the qubit   results in a time-independent Hamiltonian without the RWA (details in \aref{app:hamiltonianDerivations}),
\begin{equation}
    \frac{\hat{\tilde{H}}_\text{co}}{\hbar} = (\omega_{01} - \omega_d)\ket{1}\bra{1} - i \frac{\Omega}{2}\left(\ket{0}\bra{1} - \ket{1}\bra{0}\right).
\end{equation}

\topic{Counter-rotating drives.}
Alternatively, for $\Delta \varphi = - \pi/2$ [\fref{fig:cartoon}\panel{c}, red], the drive field travels along the equator but opposite to the qubit Larmor precession.
In a frame counter-rotating relative to the qubit at frequency $\omega_d$, we likewise find a time-independent Hamiltonian without making the RWA,
\begin{equation} % maybe delete this equation.
    \frac{\hat{\tilde{H}}_\text{counter}}{\hbar} = (\omega_{01} + \omega_d)\ket{1}\bra{1} - i \frac{\Omega}{2}\left(\ket{0}\bra{1} - \ket{1}\bra{0}\right).
\end{equation}

% \topic{Potentially redundant summary.}
% We emphasize that the realization of this mapping between circularly polarized electromagnetic fields and a fluxonium qubit drive ultimately relies on (1) charge and flux being conjugate variables and (2) time-dependent magnetic flux producing a linearly coupled flux drive. 
We emphasize that the ability to realize circularly polarized electromagnetic fields with superconducting qubit drives ultimately relies on the charge and flux drive operators being linearly independent and on the equatorial plane of the qubit Bloch sphere. 
% The requirement for distinct charge and flux drives raises a question: for a linear drive, the rotation axis of a gate defined in the rotating frame can be adjusted via the drive phase, so is it possible to create a circularly polarized drive with phase-shifted tones applied on the same control line? 
% The answer is no -- simultaneous pulses on the same control line have the same lab-frame polarization, and the counter-rotating components from both pulses cannot completely destructively interfere unless the net drive tone is zero. %From another perspective, if one tries to determine which phase-shifted pulses to apply, it also becomes apparent that this is impossible: perpendicular rotations in the rotating frame are implemented with one control line by adding a $90^\circ$ relative phase, and further phase-shifting the tones in order to create a circularly polarized drive requires another $90^\circ$ relative phase, leading to either $0$ or $180^\circ$ total relative phase. The resulting net drive is linearly polarized in either case.

\subsection{Commensurate Pulses: Regularizing Coherent Errors from Counter-Rotating Terms}\label{background:commensurate}
In this section, we detail a complementary method for mitigating counter-rotating errors even when employing a linear drive. 
For clarity, we explore resonant drives in this section, but provide analysis for the general case of off-resonant drives in \aref{app:offResonant}.
% In contrast to the implementation of co-rotating drives, this method relies solely on the pulse timing and has no additional calibration overhead.
In contrast to the implementation of co-rotating drives, this method relies solely on restricting pulse application times according to the qubit frequency, i.e., no additional calibration overhead is required relative to conventional Rabi gates.
In addition to regularizing counter-rotating effects, this approach also regularizes other sources of error, e.g., AC Stark shifts or transients which depend on pulse profiles, and can be straightforwardly implemented in parallel with typical gate calibration methods.

We introduce this method by considering a qubit subjected to a resonant, linearly polarized pulse of duration $t_g$ applied at a start time $t_0$
\begin{equation}
    \frac{\hat{H}}{\hbar} = \omega_{01} \ket{1}\bra{1} + \Omega(t - t_0) \cos(\omega_{01} t + \varphi)\left(\ket{0}\bra{1} + \ket{1}\bra{0}\right),
\end{equation}
where $\Omega(t' = t - t_0)$ denotes the envelope of the pulse such that $\Omega(t') = 0$ for $t' < 0$ and $t' > t_g$.
Rewriting the Hamiltonian in terms of $t'$ and in a frame co-rotating with the qubit at frequency $\omega_{01}$,
\begin{equation}\label{eq:rot_frame_linear_drive}
    \frac{\hat{\tilde{H}}}{\hbar} = \frac{e^{i\varphi}\Omega(t')}{2} \left[1 + e^{-2 i (\omega_{01} (t' + t_0) + \varphi)}\right] \ket{0}\bra{1} + \text{h.c.},
\end{equation}
and the time-evolution operator generated by the pulse is given by 
\begin{equation}\label{eq:pulse_unitary}
    \hat{\tilde{U}}(t'=t_g, t'=0) = \exp{\left[-\frac{i}{\hbar}\int_{0}^{t_g} \hat{\tilde{H}}(t', t_0) \,dt'\right]}.
\end{equation}
By inspecting the counter-rotating field---the second term in the square brackets of \eref{eq:rot_frame_linear_drive}---we can interpret the pulse start time $t_0$ as effectively setting a phase offset of the counter-rotating field.
When working within the RWA (neglecting the counter-rotating term), the rotating-frame Hamiltonian and resulting unitary given by \eref{eq:pulse_unitary} become invariant with respect to $t_0$.
In contrast, when applying strong linear drives with $1/t_g$ approaching $\omega_{01}/2\pi$, the counter-rotating field cannot be neglected and the unitary remains a function of $t_0$.
In other words, counter-rotating effects cause the qubit rotation to depend on when the pulse is applied~\footnote{An analogous example of physics that depends on the stability of the relative phase between the carrier and envelope of few-cycle pulses can be found in ultrafast optics~\cite{brabecIntenseFewcycleLaser2000, baltuskaAttosecondControlElectronic2003}.}. 

However, noting that this effect is coherent and deterministic, we can make the contribution from counter-rotating fields uniform for all pulses by utilizing their time periodicity.
In particular, the exponential in \eref{eq:rot_frame_linear_drive} has a period of $\tau = \pi/\omega_{01} = \tau_L/2$.
So, restricting the times at which we apply pulses to $t_0 =n\tau_L/2 + \delta t_0$ with integer $n$ and arbitrary constant time-offset $\delta t_0$ leads to the same unitary dynamics for every pulse of a fixed $\varphi$.
We set $\delta t_0 = 0$ as this time offset is equivalent to an absolute carrier-phase offset $\varphi \rightarrow \varphi -\omega_{01}\delta t_0$.
As a result of regularizing the pulse unitary, any calibration that can correct for a systematic rotation error automatically mitigates the counter-rotating contribution. 
We refer to gates performed with this approach as commensurate, since pulses are locked to a lattice defined by $\tau_L/2$.

We refer to gates not following this restriction as incommensurate and highlight their susceptibility to the coherent error generated by the non-uniform sampling of counter-rotating fields for pulses applied at different times.
We further note that when performing pulse sequences composed of a large number of incommensurate pulses, each pulse contributes a different counter-rotating error, reducing fidelity and rendering a global shift of the carrier phase $\varphi$ to all pulses indiscernable.
In contrast, a pulse sequence employing commensurate pulses will generally be sensitive to such a global offset in $\varphi$. 

To form a gate set with $X$ and $Y$ rotations (i.e. with phases $\varphi \in \{0, \pi/2\}$), we can regularize the counter-rotating dynamics for both phases simultaneously by applying $X$ gates at times $t_0^X = n\tau_L/2$ and $Y$ gates at $t_0^Y = (n + 1/2)\tau_L/2$~\cite{Campbell2020}.
Pragmatically, this can be implemented by defining $X$ gates to have a duration that is a multiple of $\tau_L/2$ and forming $Y$ gates by padding $X$ gates before and after with duration $\tau_L/4$ identity gates.
These identity gates can be compiled away for consecutive $Y$ gates to reduce the total sequence duration.
In practice, we found the benefit from this compilation to be minimal and did not utilize it for data presented in this paper.

% We now introduce a more tangible, intuitive picture of commensurate gates---that of homogenizing waveform energy in the lab frame\footnote{The interpretation of regularizing waveform power or the contribution from the counter-rotating field, the rotating-frame Hamiltonian term oscillating at the sum of the rotating frame and drive frequencies, is exactly equivalent for monochromatic drives. For polychromatic drives, in addition to regularizing counter-rotating terms, the commensurate condition must also account for interference of the drive tones. For details, see \aref{app:offResonant}.}.
% We can interpret the strong dependence of the net qubit rotation on the start time of short pulses as a consequence of the $t_0$ dependence of the total pulse energy (taking $\varphi=0$ without loss of generality), 
% \begin{align}\label{eq:power}
%     E(t_0) &\propto \int_{t_0}^{t_0 + t_g} [\Omega(t')\cos(\omega_{01} (t' + t_0))]^2 dt' \\
%     &\propto \int \Omega(t')^2[1 + \cos(2 \omega_{01} (t' + t_0))] dt'
% \end{align}
% Commensurate pulses, with $\omega_{01} t_0 / \pi \in \mathbb{Z}$, are characterized by all pulses having the same total energy in the lab frame.
% % The uniformity of the resulting waveforms is visibly apparent in \fref{fig:commensurate}\panel{e}.
% Circularly polarized drives, which have equal contributions from both quadratures but with a $\pi/2$ phase difference, do not encounter this issue as their total waveform power $\propto \Omega(t')^2 [\cos^2(\omega_{01}(t' + t_0)) + \sin^2(\omega_{01}(t' + t_0))] = \Omega(t')^2$ is inherently independent of $t_0$.

To conclude this subsection, we highlight that commensurate gates offer the ability to regularize and thereby eliminate coherent errors from the counter-rotating component of strong linear drives with no additional calibration overhead.
This stands in contrast to co-rotating drives, which natively do not include a counter-rotating component but require extra calibration.
In practice, we still found that co-rotating gates with durations approaching $\tau_L$ benefited from adhering to the commensurate condition---we attribute this to the regularization of AC Stark shifts arising from the off-resonant driving of transitions beyond $0-1$.
The AC Stark shift magnitude is dependent on when co-rotating pulses are applied due to the varying relative value of charge and flux drive components, which couple the non-target levels with different matrix elements.
This effect also grows with increasing drive strength, and its mitigation highlights another distinct benefit of commensurate gates.
We further emphasize that this approach does not rely on any specific details of the qubit.
In other words, this approach can be applied to mitigate counter-rotating effects in any platform where strong linear drives are desired.
The small price to pay for applying commensurate gates is in coherence---by requiring $X$ and $Y$ gates to be applied at times belonging to lattices shifted by $\tau_L/4$ relative to each other, the target qubit accrues decoherence during the $\tau_L/4$ waiting times between applying $X$ then $Y$ or $Y$ then $X$ gates.
For our qubit, this amounted to \SI{2.05}{ns} for every such occasion.

\subsection{Single-Qubit Gates}

In this subsection, we detail the implementation of our single-qubit gates.
We performed gates with three driving schemes: circularly polarized, purely charge~\cite{Ding2023, Somoroff2023}, and purely flux~\cite{Moskalenko2022}.

\topic{Calibrating the $\pi/2$ pulse.}
For all drive schemes, a $\pi/2$ rotation was explicitly calibrated with Rabi driving and pulse train techniques.
% For all drive schemes, a ${\pi/2}$ rotation was explicitly calibrated by first using a train of two pulses and fitting the excited state population to a cosine for the rough pulse amplitude.
% Next, an alternating pulse train was used to calibrate the pulse detuning.
% The pulse-detuning calibration can be thought of as performing a derivative removal by adiabatic gate (DRAG) calibration by only changing the drive detuning~\cite{Motzoi2009, Chow2010}.
% Then, to finely calibrate the drive amplitude, we utilized a different pulse train that yielded pseudo-identity gates with sensitivity to the pulse amplitude (e.g. only true pseudo-identities for the optimal amplitude, see \aref{app:calibration} for details).
Circularly polarized drives underwent additional calibration steps to ensure (1) equal drive strengths between the two drives, (2) a relative drive phase corresponding to a co-rotating circular polarization, and (3) simultaneity of the charge and flux pulse arrivals at the qubit.
Following the commensurate restriction required no additional calibration relative to conventional Rabi gates.
We include full details on the pulse calibration in \aref{app:calibration}.

\topic{Forming a gate set.}
With a fully calibrated $\pi/2$ rotation, we defined our $X$ ($Y$) gate as having no phase shift ($90^\circ$ phase shift) of the carrier.
% For commensurate $Y$ gates, this was accommodated by adding $\tau_L/4$ identity gates before and after each pulse, such that the counter-rotating contributions were consistent for both $X$ and $Y$ gates.
$\pi$-pulses were implemented by playing two $\pi/2$-pulses back to back, and $Z$ gates were implemented as virtual-$Z$ gates~\cite{Mckay2017}.
All microwave pulses used a pure-cosine envelope, with a \SI{0.1}{ns} gap between adjacent pulses.

\begin{figure*}[!tb]
\includegraphics {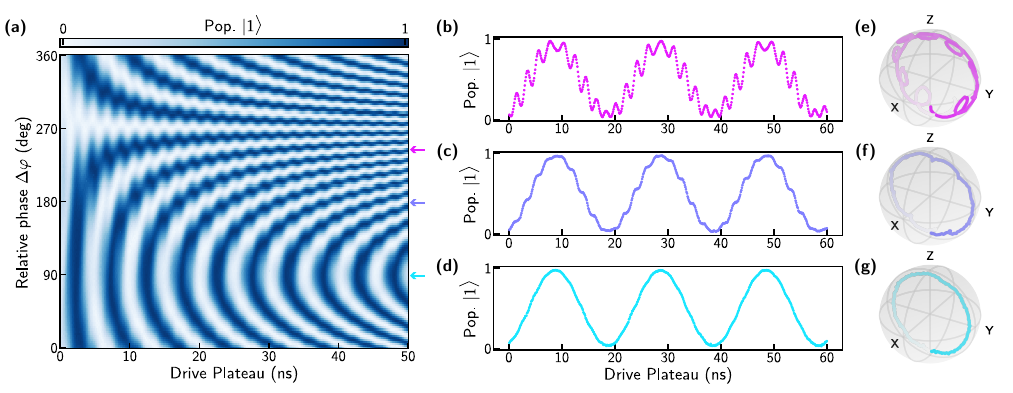}
\caption{
\label{fig:phase}
\textbf{Rabi oscillations with tunable drive polarization.}
\cpanel{a} Time-domain Rabi oscillations as the relative phase between the charge and flux drives is varied.
The relative strength of the individual charge and flux drives were calibrated to be equal, and kept constant throughout the plot.
All oscillations use a \SI{1}{ns} rise-fall time.
\cpanel{b-d} Similar data taken for three different polarizations, calibrated to each have the same Rabi frequency, all using a \SI{2.5}{ns} rise-fall time.
\cpanel{b} Nearly completely counter-rotating ($\Delta \varphi = 245^\circ$) Rabi oscillations of the fluxonium qubit.
The counter-rotating oscillations are visible on top of the slower co-rotating oscillation.
\cpanel{c} Linearly polarized drive ($\Delta\varphi = 180^\circ$) with equal contributions from charge and flux.
\cpanel{d} Completely co-rotating drive ($\Delta\varphi = 90^\circ$) illustrating elimination of counter-rotating effects.
The remaining small distortions are a result of the fast rise-time of the pulse.
\cpanel{e-f} Bloch-sphere trajectories of the corresponding oscillations (truncated to the first Rabi flop) of the data in \panel{b-d}.
Opaque color corresponds to the start time, and transparent corresponds to the end time.
}
\end{figure*}

\topic{Bencharking and gate decomposition.}
Both standard and interleaved Clifford randomized benchmarking were performed to assess the quality of our gates~\cite{Magesan2011, Magesan2012}.
We generated the single-qubit Clifford group with the gate set ${\mathcal{G} = \{I, \pm X_{\pi/2}, \pm Y_{\pi/2}\}}$, which yields on average $53/24 \approx 2.208$ non-identity gates per Clifford.
We used this decomposition as it comprises the native gates in our experiment and allows direct comparison with other recent works~\cite{liErrorSinglequbitGate2023a, hyyppaReducingLeakageSinglequbit2024} demonstrating state-of-the-art single-qubit gates.   
All fidelities were reported with uncertainties representing the standard error of the mean.

\section{Experiment}

In this section, we describe our three main results: (1) the demonstration of tunable-polarization drives, (2) the mitigation of counter-rotating errors for linearly polarized drives with commensurate gates, and (3) the optimization of gate duration for different drive schemes.

\subsection{Arbitrarily Polarized Drives}\label{exp:circ}

As our first experiment, we demonstrated the generation of arbitrarily polarized microwave drives.
To ensure equal drive strengths for the charge and flux components, we first calibrated the charge and flux drive amplitudes to individually produce the same Rabi frequency.
Then, time-domain Rabi oscillations were measured with the relative phase $\Delta \varphi$ of the two simultaneous drives swept between $0^\circ$ and $360^\circ$ [\fref{fig:phase}\panel{a}].
In this plot, the drive is shown to continuously vary between a completely co-rotating drive ($\Delta \varphi = 90^\circ$) and a completely counter-rotating drive ($\Delta \varphi = 270^\circ$).
In between, at $\Delta \varphi = 0^\circ$ and $\Delta \varphi = 180^\circ$, the drive is completely linearly polarized, but along a diagonal axis in the Bloch sphere equatorial plane [\fref{fig:cartoon}\panel{c}, blue arrows].
The counter-rotating dynamics are visibly apparent as fast oscillations on top of the slower Rabi flopping away from $\Delta \varphi = 90^\circ$.
The Rabi frequency of the simultaneous drive (within the RWA) depends on the relative phase, and is given by Eq.~\ref{eq:RabiRelativePhase}. 

% Figure 3
\begin{figure*}[!htb]
\includegraphics[width=\textwidth]{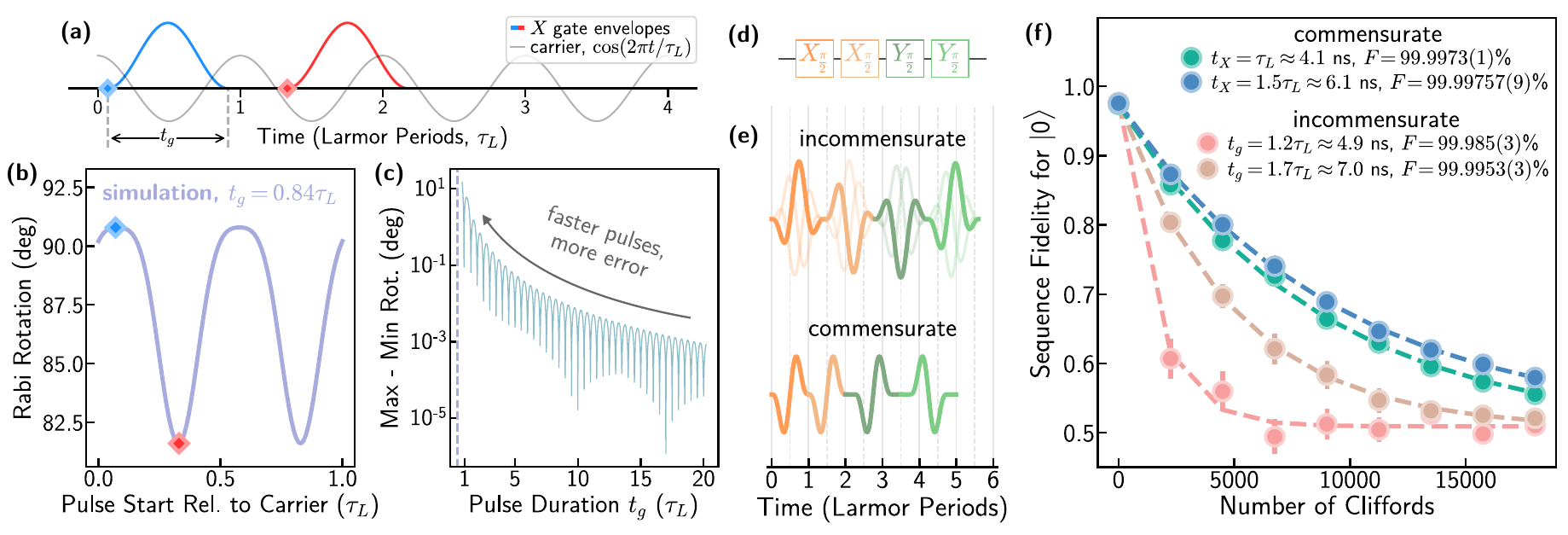}
\caption{
\label{fig:commensurate}
\textbf{Commensurate gates: alleviating counter-rotating errors for linear drives.}
\cpanel{a} Time-domain depiction of two resonant linearly polarized $X$-pulse envelopes (red and blue), starting at different times relative to the carrier ($\cos(2\pi t/\tau_L)$, grey), where $\tau_L$ is the qubit Larmor period.
\cpanel{b} Rabi rotation angle (polar angle of the Bloch vector) of an ideal two-level system starting in the ground state subject to an $X$ pulse of duration $t_g = 0.84 \tau_L$ as a function of the pulse start time modulo $\tau_L$.
The rotations from the pulses in \panel{a} are highlighted as points on the plot.
The qubit rotation depends strongly on the pulse start time due to the time dependence of the counter-rotating drive component.
\cpanel{c} Rotation angle range versus pulse duration, showing the divergence of this effect for short pulse times.
The dotted line represents the pulse duration used in \panel{b}.
% The y-axis serves as a proxy for the coherent error magnitude of this effect for gates.
\cpanel{d} Example single-qubit circuit.
\cpanel{e} Cartoon waveforms for two different implementations of the circuit drawn in \panel{d}.
Top: incommensurate pulses ($t_g = 1.2\tau_L$), which suffer from the coherent error channel depicted in panel \panel{b}.
Bottom: commensurate pulses ($t_X = \tau_L$, $t_Y = 1.5\tau_L$), which regularize the counter-rotating fields during each pulse.
This turns the coherent errors which were previously different for each pulse into a systematic coherent rotation which is corrected for automatically in our other calibrations.
For the reasons described in Section~\ref{background:commensurate}, commensurate $Y$ pulses require $\tau_L/4$ identity-gate padding before and after each pulse.
\cpanel{f} Clifford randomized benchmarking (RB) of single-qubit gates performed with flux pulses, comparing commensurate and incommensurate implementations.
All curves were averaged over 40 random seeds.
We include data from two incommensurate pulse durations ($t_g = 1.2 \tau_L \approx \SI{4.92}{ns}$ and $t_g = 1.7 \tau_L \approx \SI{6.97}{ns}$), and two commensurate implementations ($t_X = 1.0 \tau_L \approx \SI{4.1}{ns}$ and $ t_X = 1.5 \tau_L \approx \SI{6.15}{ns}$).
At these gates times, we see a significant increase in fidelity by using commensurate pulses.
This highlights the ability to mitigate coherent errors from counter-rotating terms for strong linear drives by adopting straightforward pulse-timing constraints.
}
\end{figure*}

To fairly compare the dynamics for Rabi drives of different polarizations, we then separately calibrated Rabi drives of the same strength for a mostly counter-rotating drive [\fref{fig:phase}\panel{b}], a linear drive with equal charge and flux components [\fref{fig:phase}\panel{c}], and a co-rotating drive [\fref{fig:phase}\panel{d}].
The associated Bloch sphere trajectories measured with state tomography are shown in panels \fref{fig:phase}\panel{e-g} respectively. 

In both simulation and experiment, we found that a slow rise and fall of the pulse envelope could dampen the effects of the counter-rotating terms by allowing the Hamiltonian to change adiabatically relative to the counter-rotating drive contribution~\cite{Deng2015}.
Here, to magnify the counter-rotating oscillations, we used pulses with a total rise and fall time of \SI{1}{ns} for the data in \fref{fig:phase}\panel{a} and \SI{2.5}{ns} for the data in \fref{fig:phase}\panel{b-g}.
% We attribute the slight distortions in our completely co-rotating drive \fref{fig:phase}\panel{c, g} to this fast rise/fall time and to the oscillations in the AC Stark shift caused by the varying relative strength of charge/flux drive components (which have the same period as counter-rotating oscillations, $\tau_L/4$).

The apparent tuning of the counter-rotating drive strength as a function of $\Delta \varphi$ confirms both our circuit QED analogy to optical polarization and that our two drive lines truly couple to the qubit through canonically conjugate operators; if the two drives coupled through the same operator, the Rabi frequency would still be observed to tune from a local maximum to zero as in \fref{fig:phase}\panel{a}, but no increase in counter-rotating dynamics would be observed near $\Delta \varphi = 270^\circ$.
Our technique also simplifies further research involving counter-rotating dynamics by enabling direct access to purely counter-rotating drives.

\subsection{Commensurate Gates}\label{exp:commensurate}

In this subsection, we highlight the coherent error channel posed by the non-uniform sampling of counter-rotating fields for resonant pulses applied at different times relative to the carrier phase and the mitigation of this error channel with commensurate gates. 

To establish this error channel, we simulated a two-level system subjected to an $X$ pulse applied at different start times~\fref{fig:commensurate}\panel{a}. 
We plot the the polar angle traversal of the ground state Bloch vector after application of the pulse (with duration $t_g = 0.84\tau_L$) as a function of the pulse start time in \fref{fig:commensurate}\panel{b}.
% To establish this error channel, we simulated a two-level system subject to an $X$ pulse lasting $t_G = 0.84 \tau_L$ as a function of the pulse start time [\fref{fig:commensurate}\panel{b}].
% We show two representative pulse envelopes in \fref{fig:commensurate}\panel{a}, with their rotations highlighted in \fref{fig:commensurate}\panel{b}.
% The rotation is defined as the polar angle traversal of the ground state Bloch vector after application of the pulse.
% A $0.5\tau_L$ periodicity of the effective rotation angle is observed, corresponding to the periodicity of the waveform energy.
Shifting the pulse start by $0.5 \tau_L$ corresponds to a $\pi$ phase shift of the carrier at the start of the pulse (leading to a rotation of the same amount, but in the opposite direction), resulting in a $0.5 \tau_L$ periodicity of the effective rotation angle.
% The range of the data in \fref{fig:commensurate}\panel{b} represents a proxy for the coherent error experienced by gates performed with pulses of duration $t_G = 0.84 \tau_L$.
We then simulated this effect as a function of the pulse duration [\fref{fig:commensurate}\panel{c}], plotting the rotation range as a proxy for the coherent error.
It is apparent that, as pulse durations approach $\tau_L$, this error channel grows significantly.

In order to mitigate this error channel, we adopted the commensurate approach detailed in Section~\ref{background:commensurate}.
This approach entailed applying pulses at times constrained to a lattice of spacing $\tau_L/2$ for $X$ gates (and the corresponding lattice shifted by $\tau_L/4$ for $Y$ gates, which we implemented by padding $Y$ gates before and after with $\tau_L/4$ wait times such that $Y$ gates had durations $\tau_L/2$ longer than those of $X$ gates).
Intuitively, this amounts to sampling only a single point in \fref{fig:commensurate}\panel{b}.
We draw a simple single-qubit circuit and waveforms implementing it in \fref{fig:commensurate}\panel{d,e} with both incommensurate pulses of duration $t_g = 1.2 \tau_L$ and commensurate pulses of duration $t_X = \tau_L$ ($t_Y = 1.5\tau_L$).

We implemented both commensurate and incommensurate single-qubit gates and characterized their performance with Clifford randomized benchmarking [\fref{fig:commensurate}\panel{f}].
% For commensurate gates, there is an additional degree of freedom corresponding to the absolute phase offset of the carrier, which effectively determines the shape of the regularized waveform (for incommensurate gates, this phase is inconsequential since different pulses sample different carrier phases).
For commensurate gates, we additionally optimized the absolute carrier phase offset,  which effectively determines the shape of the regularized waveform. The dependence of the fidelity on this phase was larger for shorter gates, with a most extreme variation of $5 \times 10^{-5}$.
% We empirically optimized the absolute carrier phase offset for commensurate gates and generally found a larger variation in the fidelity for shorter gates with a most extreme variation of $5 \times 10^{-5}$.
We attribute this dependence to the compensation of, e.g., transients or a slight offset of the qubit flux bias, which would affect the qubit trajectory through the 0-1 avoided crossing.
Notably, commensurate gates with durations as low as $\tau_L$ were characterized by fidelities well in excess of $99.997\%$, in contrast to incommensurate pulses at similar times which were characterized by fidelities as low as $99.985\%$.
As expected, the benefit from commensurate pulses was more significant for faster gates.

\subsection{Linear and Circular Gate Characterization}
\begin{figure}[!htb]
\includegraphics[width=0.45\textwidth]{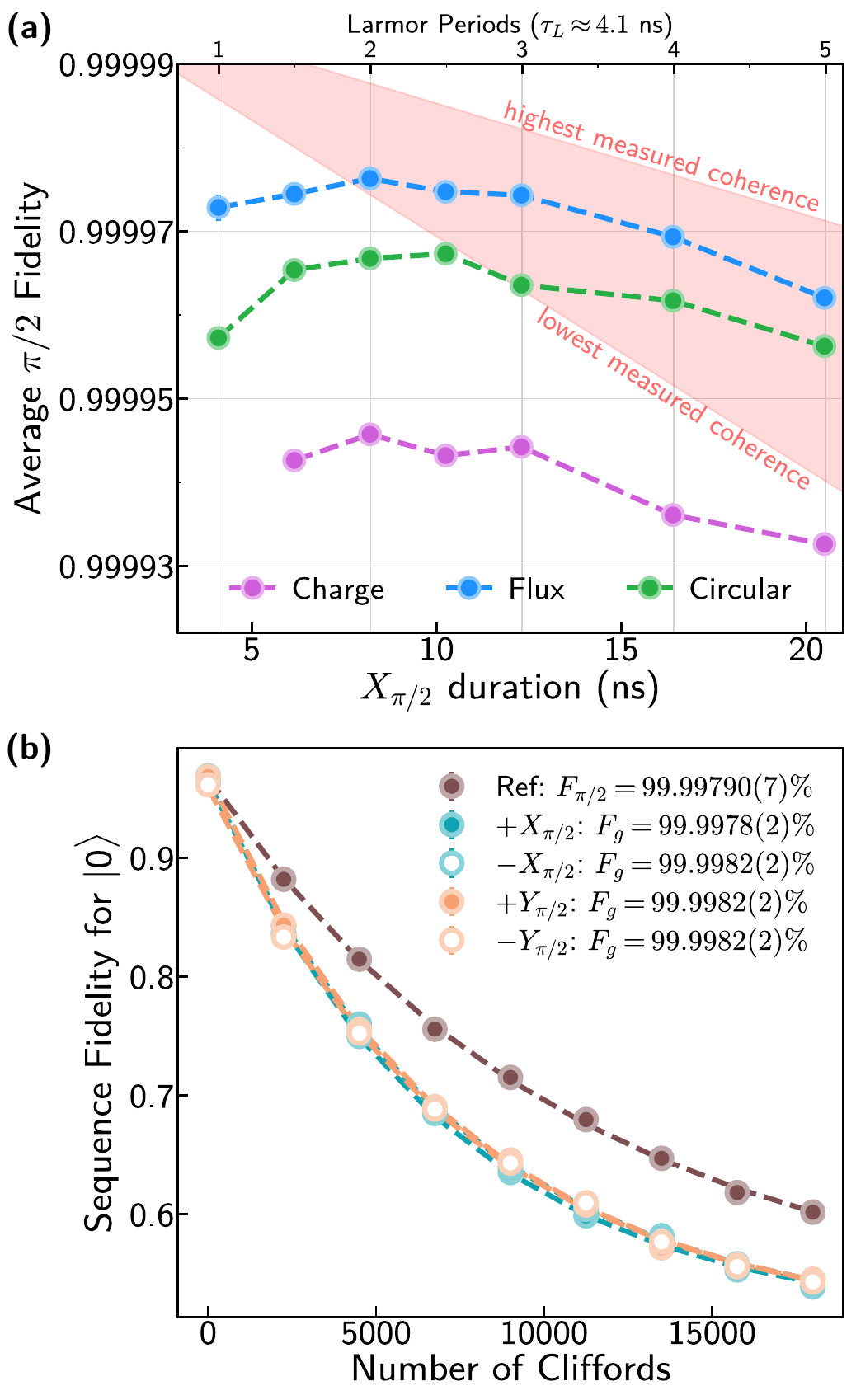}
\caption{
\label{fig:fidelities}
\textbf{Commensurate single-qubit gates with linear and circularly polarized drives.}
\cpanel{a} Gate fidelity measured with Clifford randomized benchmarking as a function of the $\pi/2$-pulse width for a pure charge drive (purple), pure flux drive (blue), and a co-rotating circularly polarized drive (green).
In order to mitigate the effect of coherence fluctuations, data was collected over a span of two weeks with intermittent breaks taken when the qubit coherence fluctuated to low values.
The lowest and highest measured coherences during this window ($\SI{300}{\micro s} \lesssim T_1 \lesssim \SI{500}{\micro s}$, $\SI{200}{\micro s} \lesssim T_{2E} \lesssim \SI{500}{\micro s}$) were used to calculate fidelity bounds (red).
\cpanel{b} Clifford interleaved randomized benchmarking for a calibrated linearly polarized flux drive with an $X_{\pi/2}$ duration of $t_X = 2\tau_L$.
The reference trace has an average single-qubit gate error of $(2.10 \pm 0.07) \times 10^{-5}$.
All randomized benchmarking traces were averaged over 25 random seeds.
}
\end{figure}

% \topic{Gates vs. width.}
As a final experiment, we empirically optimized the duration of our commensurate gates by measuring the average single-qubit gate fidelity as a function of the gate duration for three driving schemes: charge, flux, and co-rotating drives [\fref{fig:fidelities}\panel{a}].
Coherence bounds were calculated using the average gate time $t_g = (t_X + t_Y)/2 = t_X + \tau_L/4$.
% We found optimal gate fidelities for all drive schemes near $t_X = 2\tau_L \approx \SI{8.2}{ns}$.
We found all drive schemes were optimal near $t_X = 2\tau_L \approx \SI{8.2}{ns}$, with fidelities $>99.994\%$ for charge, $>99.996\%$ for circular co-rotating, and $>99.997\%$ for flux drives. 
Linear flux (charge) drives resulted in our best (worst) gates, with circular drives performing better than charge but worse than flux.
We separately performed interleaved randomized benchmarking using the $t_X = 2\tau_L$ flux pulses for each gate in our gate set [\fref{fig:fidelities}\panel{b}]. 
% Notably, we found that charge driven gates had consistently lower fidelities than flux and circular gates, which was replicated in numerical simulations. 
We further explored the distribution of errors and gate stability in \aref{app:errors}.
We present additional data comparing incommensurate linear and circular gates in \aref{app:incomm_flux_vs_circ}, demonstrating that circular drives can benefit incommensurate gates, but do not outperform flux-driven commensurate gates.

% For our qubit parameters, since the charge operator couples non-computational state transitions (notably $\ket{1}\leftrightarrow \ket{2}$ and $\ket{0}\leftrightarrow \ket{3}$) more strongly than the flux operator for a fixed gate time, we hypothesize that this difference in fidelity was caused by leakage into higher levels. 

% Figure 4

\section{Outlook}

The effect of counter-rotating dynamics on gate operations has long been an understudied aspect of Rabi-oscillation-based gates.
With the growing popularity of fluxonium and other low-frequency qubits as well as the ongoing development of control strategies to mitigate other dominant error channels~\cite{hyyppaReducingLeakageSinglequbit2024}, counter-rotating effects have become increasingly relevant for calibrating fast, high-fidelity microwave gates.
We introduced and demonstrated two complementary approaches for mitigating these effects. 

Our first protocol uses independently controlled linear charge and flux drives, representing two independent quadratures in circuit QED.
This control enables the ability to change the polarization angle of the total effective drive, giving a controllable means to study counter-rotating dynamics and the trade-offs between different linear drives.
Using such control, we demonstrated the ability to create a 
circularly polarized drive---natively lacking a counter-rotating component, independent of the drive amplitude---for use in gates.

Our second protocol, commensurate pulses, leverages the coherent and time-periodic nature of counter-rotating fields to regularize their contributions to all control pulses, thereby eliminating the counter-rotating error channel for linear drives.
% This protocol relies on the application of gates at discrete times belonging to lattices defined by half the qubit Larmor period, is platform independent, and requires no calibration overhead.
This protocol relies on the application of gates at periodically discrete times determined by the qubit Larmor period, is platform independent, and requires no additional calibration overhead for Rabi-based gates.
We established the efficacy of this protocol for gates as short as one Larmor period, finding almost an order of magnitude improvement in fidelity over incommensurate pulses of a similar duration. 

We note that our commensurate pulse technique shares close resemblance with those used for Landau-Zener-based single-qubit gates~\cite{Campbell2020, Zhang2021}.
% ~\footnotetext[2]{We also note a related example of coherent population transfer dependent on a commensuration of drive amplitude and frequency (as opposed to drive frequency and pulse application time in this work) in the context of multi-pass Landau-Zener-Stuckelberg interferometry~\cite{Oliver2005, Berns2006, Rudner2008, Berns2008}.}.
In both schemes, the exact specification of the waveform, including its timing, is crucial for the fidelity of the gate.
% One crucial difference is the reference frame in which the gates are defined; commensurate pulses utilize a rotating frame, whereas Landau-Zener gates do not.
However, these gates are often viewed from a different perspective; commensurate pulses are readily modeled in a rotating frame, whereas Landau-Zener gates are not.
In the lab frame, the natural qubit precession supports $Z_{\phi(t)}$ gates by idling for a time $t$.
Identity gates that are not exact multiples of the Larmor period must then be performed by modulating the qubit frequency~\cite{Weiss2022}.
Likewise, identity gates of arbitrary duration cannot be performed in the rotating frame without relaxing the commensurate pulse conditions, and $Z$~gates are implemented via a combination of idling and phase shifting the next drive pulse so that it is once again commensurate.
The main advantage of commensurate pulses is the ability to use a stable underlying carrier frequency in hardware (e.g. with a dedicated RF source). With this in place, errors in envelope timing only translate into small errors in the polar rotation angle [\fref{fig:commensurate}\panel{c}], whereas without, errors in timing freely rotate the equatorial axis of the gate. This advantage is more pronounced when the qubit frequency is higher.
While the language to describe these operations is often different, both pairs of gates rely on the same underlying physics of the natural qubit precession to perform $I$- and $Z$-gates. 

We demonstrated state-of-the-art single-qubit gates with three separate drive schemes on the same qubit (pure charge, pure flux, and circularly polarized), and found that flux pulses performed the best.
We attribute this to extra decoherence from system heating associated with the charge drive (see \aref{app:errors}) and the matrix element structure of the fluxonium at half-flux (featuring larger charge matrix elements for transitions to states outside the computational subspace).
We hypothesize that, if leakage transitions and heating were mitigated for charge drives, circularly polarized drives may potentially benefit Rabi-based gates for even more aggressive gate times.
Furthermore, a circularly polarized drive remains an attractive tool when optimizing single-qubit gates in even lower-frequency qubits~\cite{Zhang2021, Earnest2018, Ficheux2021, Campbell2020} and for other low-frequency microwave interactions such as cross-resonance~\cite{Dogan2023}, iSWAP~\cite{Zhang2024}, or bSWAP gates~\cite{Nesterov2021}. 

We expect commensurate pulses, being platform-independent and requiring no additional calibration overhead, to benefit any platform where fast, resonant control is desired and counter-rotating dynamics are problematic.
% For two-qubit gates based on resonant control, the same arguments for homogenizing the counter-rotating dynamics can be applied within the computational-state subspace for the gate, and the relevant lattice for pulse start times will be defined analogously by the relevant carrier frequency (e.g. the transition frequency for a MAP gate).
% The timing constraints for the commensurate implementation of a multi-qubit circuit will demand the careful orchestration of pulses in the time domain, but the only price to pay (beyond implementation) will be the insertion of idle time to accommodate Z-rotations and the synchronization of multiple qubits.
Although our demonstration only included one qubit, we note the straightforward extension of commensurate pulses to a multi-qubit processor based on resonant single-qubit control.
Each qubit $i$ will have an associated Larmor period $\tau_L^i$, and so control pulses for each qubit will follow their individual lattices as defined in \ref{background:commensurate}.
Any multi-qubit gate then simply needs to allocate a requisite number of Larmor periods from each qubit to accommodate the duration of the added gate. Aside from the short idling times this requires, the compilation of commensurate pulses with multiple qubits needs no other special consideration.

{\em Note}: While preparing this manuscript, we became aware of a related work~\cite{Sank2024}%~\footnote{Private communication, Daniel Sank.}

\begin{acknowledgments}
We gratefully acknowledge insightful conversations with R\'eouven Assouly, Samuel Alipour-Fard, David Newsome, Lamia Ateshian, Amir Karamlou, Youngkyu Sung, and Agustin Di Paolo.
This research was funded in part by the U.S. Army Research Office under Award No. W911NF-23-1-0045; in part by the U.S. Department of Energy, Office of Science, National Quantum Information Science Research Centers, Co-design Center for Quantum Advantage (C2QA) under contract number DE-SC0012704; and in part under Air Force Contract No. FA8702-15-D-0001.
M.H., P.M.H., and I.T.R. are supported by an appointment to the Intelligence Community Postdoctoral Research Fellowship Program at MIT administered by Oak Ridge Institute for Science and Education (ORISE) through an interagency agreement between the U.S. Department of Energy and the Office of the Director of National Intelligence (ODNI).
 D.A.R. gratefully acknowledges support from the NSF (award DMR-1747426).
 Any opinions, findings, conclusions or recommendations expressed in this material are those of the author(s) and do not necessarily reflect the views of the US Air Force or the US Government.

% D.A.R and L.D. performed the experiments and analyzed the data.
% L.D., M.H., and J.A. developed the theory and numerical simulations.
% L.D., Y.S., A.D.P., and K.S. designed the device.
% K.A., D.K.K., B.M.N., A.M., M.E.S., and J.L.Y. fabricated and packaged the devices. 
% L.D., M.H., Y.S., B.K., J.A., A.H.K., and S.G. contributed to the experimental setup.
% T.H., T.P.O., S.G., J.A.G., K.S., and W.D.O supervised the project.
% L.D. and M.H. wrote the manuscript with feedback from all authors. 
% All authors contributed to the discussion of the results and the manuscript. 

\end{acknowledgments}

\appendix

\section{Wiring and Control Hardware}\label{appendix:wiring}
\begin{figure}[!htb]
\includegraphics{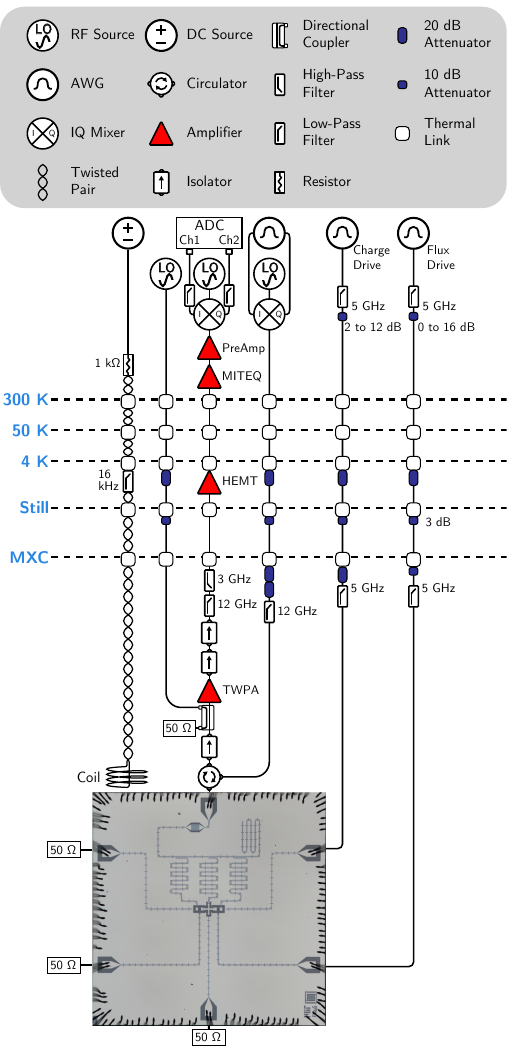}
\caption{\label{fig:sup_wiring} \textbf{Wiring schematic of the experimental setup.}} 
\end{figure}

All experiments were conducted in a Bluefors XLD600 dilution refrigerator maintaining a base temperature stabilized at $\sim \SI{22}{mK}$.
Flux biases were provided by a small superconducting solenoid mounted to the lid of the sample package.
We specify the equipment used for qubit biasing, control, and readout in \tref{tab:equipment}, and detail the experimental wiring in \fref{fig:sup_wiring}.

\begin{table}[!htb]
\caption{\label{tab:equipment}
\textbf{Summary of control equipment.}
}
\begin{tabular}{|c|c|c|}
\hline
Component & Manufacturer & Model\\
\hline
Dilution Refrigerator & Bluefors & XLD600\\
RF Source & Rohde and Schwarz & SGS100A\\
DC Source & QDevil & QDAC I\\
Control Chassis & Keysight & M9019A\\
AWG (readout pulses) & Keysight & M3202A\\
AWG (qubit pulses) & Keysight & M8195A\\
ADC & Keysight & M3102A\\
\hline
\end{tabular}
\end{table}

\section{Circularly Polarized Driving}\label{app:hamiltonianDerivations}
In this section, we present derivations for the co- and counter-rotating drive Hamiltonians in the fluxonium qubit subspace. We also derive an expression for the Rabi frequency (within the RWA) as a function of the relative phase between simultaneous charge and flux drives.

Truncating to the ground and first excited state manifold of the system Hamiltonian Eq.~\ref{eq:fullHamiltonian}, the two-level Hamiltonian terms take the following form 
\begin{align*}
    \hat{H}_0 &\rightarrow \hbar \omega_{01} \ket{1}\bra{1}, \\
    \hat{n} &\rightarrow i(\ket{0}\bra{1} - \ket{1}\bra{0}), \\
    \hat{\phi} &\rightarrow \ket{0}\bra{1} + \ket{1}\bra{0}.
\end{align*}

Considering simultaneous charge and flux drives of the same strength ${\Omega_c \bra{0}\hat{n}\ket{1} = \Omega_f \bra{0}\hat{\phi}\ket{1} \equiv \Omega / 2}$ then yields the two-level system Hamiltonian
\begin{align*}
    \frac{\hat{H}}{\hbar} = \;\, &\omega_{01} \ket{1}\bra{1} \\
    &- i\frac{\Omega}{4} \cos(\omega_d t)(\ket{0}\bra{1} - \ket{1}\bra{0}) \\
    &+ \frac{\Omega}{4}\cos(\omega_d t - \Delta \varphi)(\ket{0}\bra{1} + \ket{1}\bra{0}).
\end{align*}

\subsection{Co-rotating Drive}

Applying a rotating frame transformation
\begin{equation}
    \hat{\tilde{H}} = \hat{U} \hat{H} \hat{U}^{-1} + i \hbar \frac{\partial \hat{U}}{\partial t} \hat{U}^{-1}
\end{equation}
at a frequency $\omega_d$ co-rotating with the qubit, $\hat{U}_\text{rf}(t) = e^{i \omega_d t \ket{1}\bra{1}}$ gives
\begin{align}
    \frac{\hat{\tilde{H}}}{\hbar} = \,\,&(\omega_{01} - \omega_d) \ket{1}\bra{1} \nonumber\\
        &-i \frac{\Omega}{4}\left[1 + ie^{-i \Delta \varphi} + e^{-2i\omega_d t}(1 + ie^{i\Delta \varphi})\right]\ket{0}\bra{1} + \text{h.c.}\label{eq:twoLevelRotFrameHamiltonian}
\end{align}
For $\Delta \varphi = \pi/2$, the fast-oscillating terms exactly cancel, and the Hamiltonian simplifies to a qubit with a static drive field (without needing the RWA),
\begin{equation}
    \frac{\hat{\tilde{H}}_\text{co}}{\hbar} = (\omega_{01} - \omega_d) \ket{1}\bra{1} -i \frac{\Omega}{2} (\ket{0}\bra{1} - \ket{1}\bra{0}).
\end{equation}

\subsection{Counter-rotating Drive}

For $\Delta \varphi = -\pi/2$, following a similar procedure with a rotating-frame transformation in the opposite direction, $\hat{U}_\text{rf}(t) = e^{-i \omega_d t \ket{1}\bra{1}}$, gives
\begin{equation}
    \frac{\hat{\tilde{H}}_\text{counter}}{\hbar} = (\omega_{01} + \omega_d) \ket{1}\bra{1} -i \frac{\Omega}{2} (\ket{0}\bra{1} - \ket{1}\bra{0}).
\end{equation}

\subsection{Generalized Rabi Oscillations with Simultaneous Drives of Arbitrary Relative Phase}
With resonant drives ($\omega_d = \omega_{01}$), the rotating frame Hamiltonian \eref{eq:twoLevelRotFrameHamiltonian} only retains the coupling term
\begin{equation}
    \frac{\hat{\tilde{H}}}{\hbar} = -i \frac{\Omega}{4}(1 + ie^{-i \Delta \varphi} + e^{-2i\omega_d t}(1 + ie^{i\Delta \varphi}))\ket{0}\bra{1} + \text{h.c.}
\end{equation}
% \begin{equation}
%     \frac{H_\text{rot}}{\hbar} = \begin{pmatrix}
%                     0 & \frac{\Omega}{2}[1 - ie^{i\Delta\varphi} + e^{-2 i\Delta\varphi\omega_{d}t} (1 - ie^{-i \Delta \varphi})] \\
%                     h.c. & 0 
%                     \end{pmatrix}.
% \end{equation}
For weak drives ($\Omega \ll \omega_{01}$), we can perform the RWA (discard terms oscillating at $2 \omega_d$) and the qubit will display generalized Rabi oscillations at a frequency
\begin{align}\label{eq:RabiRelativePhase}
    \tilde{\Omega} &= \frac{\Omega}{4} |1 + ie^{-i\Delta \varphi} | \\
                &= \frac{\Omega}{2} |\cos((\Delta \varphi - \pi/2)/2)| \\
                &= \frac{\Omega}{2} \sqrt{\frac{1 + \cos(\Delta \varphi - \pi/2)}{2}}.\label{eq:rabi_vs_relative_phase}
\end{align}
To determine the relative phase offset of the charge and flux drives, we applied simultaneous drives of the same strength while sweeping the relative phase and fit the observed oscillation frequencies to \eref{eq:rabi_vs_relative_phase}.

\section{Commensurate Condition for Off-Resonant Pulses}\label{app:offResonant}
In this section, we derive a commensurate condition ensuring the uniformity of the qubit rotation for different pulse start times utilizing an off-resonant drive.

Consider a qubit subjected to a linearly polarized pulse of duration $t_g$ with carrier frequency $\omega_{d} = \omega_{01} + \delta$ applied at a variable start time $t_0$
\begin{equation}
    \frac{\hat{H}}{\hbar} = \omega_{01} \ket{1}\bra{1} + \Omega(t - t_0)\cos(\omega_{d} t + \varphi)\ket{0}\bra{1} + \text{h.c.}
\end{equation}
where $\Omega(t' = t - t_0)$ denotes the pulse envelope such that $\Omega(t') = 0$ for $t' < 0$ and $t' > t_g$.
Rewriting the Hamiltonian in terms of $t'$ and performing the standard rotating-frame transformation co-rotating with the qubit at frequency $\omega_{01}$, we obtain
\begin{equation}\label{eq:rot_frame_detuned}
   \frac{\hat{\tilde{H}}}{\hbar} = e^{i (\varphi + \delta t_0)} \frac{\Omega(t')}{2} \left[e^{i\delta t'} + e^{-i(\omega_{\Sigma} t' + 2\omega_{d}t_0 +2\varphi)} \right]\ket{0}\bra{1} + \text{h.c.}
\end{equation}
with $\omega_{\Sigma} \equiv \omega_{d} + \omega_{01}$.
The time-evolution operator generated by the pulse is given by 
\begin{equation}\label{eq:poly_unitary}
    \hat{\tilde{U}}(t'=t_g, t'=0) = \exp{\left[-\frac{i}{\hbar}\int_{0}^{t_g} \hat{\tilde{H}}(t', t_0) \,dt'\right]}.
\end{equation}

\topic{Monochromatic, resonant drive.} In this case, ${\delta = 0}$ and \eref{eq:rot_frame_detuned} reduces to \eref{eq:rot_frame_linear_drive}.
To regularize the unitary for all pulse start times $t_0$ of a given $\varphi$, we require $t_0 = n\tau_L/2 + \Delta t_0$ with integer $n$ and arbitrary constant time-offset $\Delta t_0$.

\topic{Monochromatic, off-resonant drive.} In this case, $\delta \neq 0$.
In contrast to the resonant case, we now have an additional $t_0$ dependence in the phase prefactor in \eref{eq:rot_frame_detuned} which affects the axis of rotation.
However, this prefactor is a constant with respect to the $t'$ integration of \eref{eq:poly_unitary}.
To regularize the unitary for all pulse start times $t_0$, it suffices to regularize the counter-rotating term with $t_0 = n\tau_d/2 + \Delta t_0$, where $\tau_d = 2\pi/\omega_d$ is the drive period.
The axis of rotation can be corrected for with a subsequent $Z$ rotation.
% We can equivalently interpret this commensurate condition as regularizing the waveform energy, $E(t_0) \propto \int \Omega(t')^2[1 + \cos(2 \omega_d (t' + t_0))] dt'$, which is periodic in $t_0$ with period $\tau_d/2$.

\section{Single-Qubit Gate Calibration}\label{app:calibration}
% potential comment on all pi/2: We note that for a given pulse sequence, this choice results in a longer sequence duration and therefore a more severe coherence limit to gate fidelities.
All gates in this work were made of $\pi/2$-pulses (e.g. $\pi$-pulses utilized two sequential $\pi/2$-pulses) with cosine envelopes and performed at $\Phi_{\text{ext}} = 0.5 \Phi_0$, referred to as the ``sweet spot."
A pulse starting at time $t_0$ with duration $t_g$ was implemented with the waveform $w(t)$,
\begin{align*}
    w(t) &= A(t) \cos[\omega_{01} t - \delta (t - t_0) + \phi] \\ 
    A(t) &= 
    \begin{cases} 
        \frac{A_0}{2} (1 - \cos[2\pi \frac{(t - t_0)}{t_g}])\text{, } &t_0 \leq t \leq t_0 + t_g \\
        0\text{, } & \text{otherwise}
    \end{cases} 
\end{align*}
where $\omega_{01}/2\pi$ is the undriven qubit frequency, $\delta$ is the pulse detuning, and $\phi$ is the phase of the carrier defining the axis of the qubit rotation.

Before precise gate calibration, we performed rough calibrations of the flux bias, qubit frequency, and a slow ($\geq 30$ ns) charge $\pi/2$-pulse for precise flux bias and frequency calibrations.
The following preliminary calibrations were performed before gate tune-up.

\begin{enumerate}
    \item \textbf{Precise flux bias calibration.} Ramsey oscillations were measured with a fixed pulse frequency (slightly detuned below the qubit frequency at the sweet spot) as a function of flux bias.
    The oscillation frequencies vs flux were fit to a parabola, with a minimum oscillation frequency at the flux sweet spot.
    For details, we refer to~\cite{Ding2023}.
    \item \textbf{Single-shot readout calibration.} We collected single-shot voltage measurements with no initialization of the qubit, and trained a Gaussian mixture model on the resulting dataset in the IQ plane.
    \item \textbf{Precise qubit frequency calibration.}
    Ramsey oscillations were measured for \SI{10}{\micro s} with an applied detuning of \SI{-800}{kHz} and fit with a cosine function.
    The difference between the fit oscillation frequency and the applied detuning was then used to calculate the precise qubit frequency.
\end{enumerate}

Linear drive gates for a given $\pi/2$-pulse duration were calibrated roughly following the procedure of \cite{Ding2023}, with minor adjustments.
We detail the exact procedure below. 
\begin{enumerate}
    \item \textbf{Rough $\pi$-pulse amplitude calibration.}
    A train of two Rabi pulses was applied to the qubit while sweeping the amplitude of their cosine envelopes.
    The resulting population of the excited state was fit with a cosine function, yielding the rough amplitude corresponding to two sequential $\pi/2$ pulses.
    \item \textbf{Pulse detuning.}
    A train of $\pi/2$ pulses with alternating signs (e.g. $[X_{\pi/2} + X_{-\pi/2}]^n$) was applied while sweeping the number of pulses and a detuning modulation of the envelope of each pulse given by $e^{-i \delta (t -t_0)}$, where $t_0$ was the start time of each pulse.
    The optimal value of $\delta$ was chosen such that the oscillation between $\ket{0}$ and $\ket{1}$ was minimized.
    % Experiments showed that optimizing this modulation frequency was equivalent in our setup to optimizing the DRAG parameter, suggesting that this calibration compensated for errors arising from e.g. AC-stark shifts from higher levels, the Bloch-Siegert shift, or line distortions, rather than leakage.
    Interestingly, for commensurate gates, we found that the particular pulse used for this calibration (e.g. $X$ or $Y$) was important -- for gate durations $t_X = n \tau_L$ ($t_X = (n + 1/2) \tau_L$), $n\in\mathbb{N}$, we found that this pulse sequence behaved as expected when using $Y$ ($X$) rotations.
    Recall that for our commensurate implementation, $Y$ gates have a duration $\tau_L/2$ longer than $X$ gates.
    This results in, for $t_X = n \tau_L$ ($t_X = (n + 1/2) \tau_L$), alternating $Y$ ($X$) pulse trains with identical waveforms for all gates (in contrast, the equivalent same-sign pulse trains correspond to the waveform for each subsequent gate being flipped).
    We hypothesize that, due to sequential pulses implementing rotations in opposite directions having the same waveform, long-time transients were suppressed rather than amplified in this calibration by the choice of $X$ vs $Y$ rotations.
    \item \textbf{Precise amplitude calibration. \label{calib:LinearPreciseAmp}}
    We used a pulse train looping over the set $\{X_{\pi/2}, Y_{\pi/2}, X_{-\pi/2}, Y_{-\pi/2}\}$ measuring only trains with a multiple of 3 pulses (e.g. $n_\text{tot} = 3 n$, $n\in\mathbb{N}$).
    This sequence implements a pseudo-identity gate (only an identity with perfect control) that is sensitive to the pulse amplitude.
    We also note that this sequence was sensitive to the pulse detuning.
    However, the pulse detuning was calibrated immediately preceding and no effective change was seen by changing this calibration to the more usual $X_\pi^n$ train.
    This specific pulse sequence was chosen in order to treat $X$ and $Y$ pulses equally and mitigate any skewing of the optimal amplitude from e.g. unwanted amplification of transients from a homogeneous pulse train.
\end{enumerate}

% talk about why this order / what affected what?
For circular gates, three additional calibrations were performed to determine 1) the relative phase offset, 2) the relative pulse delay, and 3) the precise relative amplitude between the charge and flux drives.
The exact procedure is outlined below.
\begin{enumerate}
    \item \textbf{Rough $\pi$-pulse amplitude calibration for charge and flux.}
    We applied a sequence $X_{\pi} + X_{\pi}$ while sweeping the pulse amplitude and fit the resulting population of $\ket{1}$ to a cosine function. The optimal $\pi$-pulse amplitude was then calculated as half the amplitude for the max population. 
    \item \textbf{Relative phase offset calibration.}
    We applied simultaneous charge and flux drives of equal strength (with relative amplitude set by the previous measurement) and measured Rabi oscillations while sweeping the relative phase. The offset phase was determined by fitting the data to \eref{eq:RabiRelativePhase}.
    \item \textbf{Relative pulse delay calibration.}
    We applied an alternating $X_{\pi}$ pulse train of nominally simultaneous charge and flux gates while sweeping the relative drive delay and the total number of pulses (constraining it to be odd) and setting $\Delta \varphi = 0$.
    The resulting interference pattern yielded maximal population of $\ket{1}$ when the pulse trains for charge and flux arrived at the qubit simultaneously.
    \item \textbf{Pulse detuning calibration (coarse). \label{calib:CircDrag}}
    The optimal pulse detuning parameter for circular gates was found with the same routine as for linear gates, with the relative phase set to $\Delta \varphi = \pi/2$ and enforcing  $\delta = \delta_\text{charge} = \delta_\text{flux}$. 
    \item \textbf{Relative amplitude calibration.} We applied a nominally counter-rotating Rabi drive ($\Delta\varphi = -\pi/2$) with a plateau equal to 3 times the pulse width as a function of the flux pulse amplitude (while keeping the charge amplitude fixed), resulting in minimal excited state occupation when the Rabi strengths of charge and flux were balanced.
    \item \textbf{Pulse detuning calibration (fine).} As described in step \ref{calib:CircDrag}.
    \item \textbf{Precise amplitude calibration.}  As described in the linear gate calibration, step \ref{calib:LinearPreciseAmp}.
\end{enumerate}

% Use PNG later on to get rid of the white square borders
\begin{figure*}[!tb]
\includegraphics{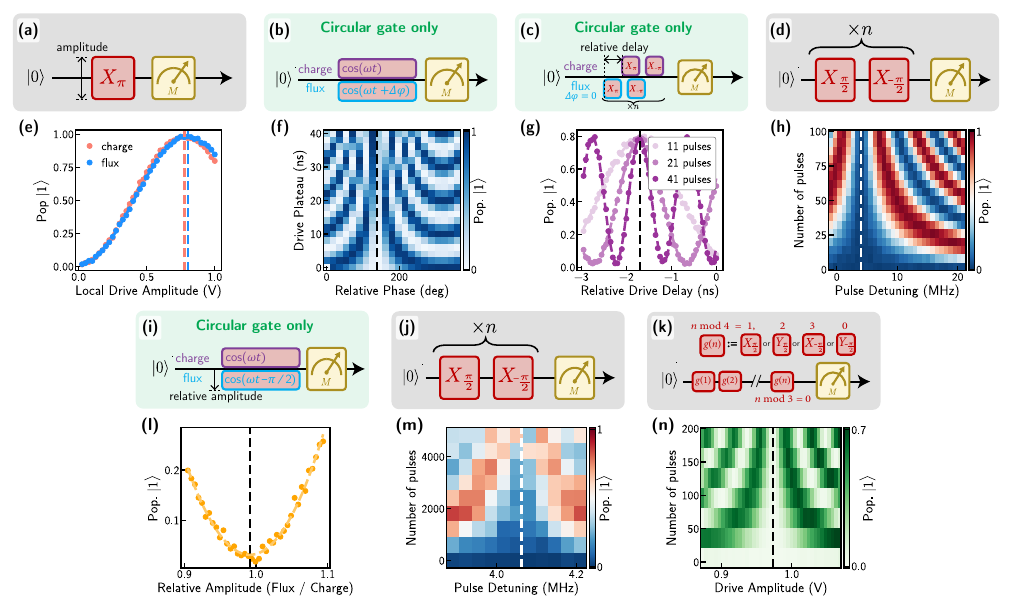}
\caption{
\label{fig:sup_calibration}
\textbf{Measurement pulse sequences for single-qubit gate calibration.}
\cpanel{a-d},\cpanel{i-k} Measurement pulse sequences for \cpanel{e-h},\cpanel{l-n}, respectively.
\cpanel{e} Rough pulse amplitude calibration.
\cpanel{f} Circular-only calibration for the relative phase between charge and flux drives. 
Marked in black is the counter-rotating relative phase.
All circular gates were performed with the co-rotating phase ($180^\circ$ offset from the counter-rotating phase).
\cpanel{g} Circular-only calibration for the relative delay between charge and flux pulses, to ensure pulses sent down both control lines arrived simultaneously at the qubit.
\cpanel{h} Rough pulse detuning calibration, primarily compensating for AC Stark shifts.
\cpanel{l} Circular-only fine calibration to balance the charge and flux drive strengths.
\cpanel{m} Fine pulse detuning calibration.
\cpanel{n} Fine pulse amplitude calibration.}
\end{figure*}

\section{Gaussian Fitting of Calibration Scans: \\Gaussian Function as an Infinite Product}

For multiple pulse parameters, optimal values were determined by fitting the product of pulse-train dataset line-cuts to a Gaussian function (in other words, multiplying the horizontal rows of a dataset which looks like \fref{fig:sup_calibration}\panel{h}, and fitting the resulting line).
In this subsection, we justify this methodology and provide an infinite product expression of the Gaussian function $e^{-x^2 / 2\sigma^2}$.

For a pulse parameter $x$, we assume that at the optimal value $x_\text{opt}$ applying a pseudo-identity gate comprised of a pulse train with $n$ repetitions (e.g. $(X_{\pi}+X_{-\pi})^n$) leaves the qubit in its initial state.
However, a slightly offset value of $x$ will lead to an over- or under-rotation of the qubit by an amount $n \Delta\theta = n k (x-x_\text{opt})$ and a corresponding population of the excited state of $p_{\ket{0}}^n = \frac{1}{2} (1 + \cos(n \Delta \theta))$, where $k$ is a constant of proportionality depending on the details of how the parameter $x$ affects the qubit rotation.
In our experiment, $x$ corresponded to the pulse detuning or amplitude, but we note that this argument can be generalized to other parameters which effect the qubit rotation angle similarly.

In order to determine the optimal value $x_\text{opt}$, we took datasets measuring the excited state probability $p_{\ket{1}}^n$ after pseudo-identity gates as a function of the number of repetitions $n$ and the pulse parameter $x$.
We then took the product for all different $n$, resulting in the signal $s(x)$, 
\begin{align}
    s(x) &= \prod_{n=0}^{N} \frac{1}{2}(1+\cos(n \Delta \theta)) \\
         &= \left(\prod_{n=0}^{N} \cos(n \Delta \theta)\right)^2.
\end{align}
In the limit as $N\to\infty$, one can show that such an infinite product converges exactly to a Gaussian when the cosine frequencies are weighted correctly, 
\begin{align}
    \lim_{N\to\infty} \prod_{n=0}^{N} \cos\left(\frac{n}{N^{3/2}} \Delta \theta\right) = e^{-\Delta \theta^2 / 6}.
\end{align}

Even for small $N$ (in our datasets, $N\leq 5$), the above expression is easily fit to a Gaussian, giving an efficient and robust way to fit datasets such as \fref{fig:sup_calibration}\panel{g,h,m,n}.

\section{Gate Errors \& Stability}\label{app:errors}

\subsection{Coherent and Incoherent Errors}\label{app:PRB}

To quantify the proportion of incoherent and coherent error in our gates, we performed purity randomized benchmarking (RB)~\cite{wallmanEstimatingCoherenceNoise2015a, fengEstimatingCoherenceNoise2016}.
Our implementation of the pulse sequence included a recovery gate as in RB, followed by single-qubit state tomography.
This enabled us to extract both the total and incoherent error per gate by analyzing $\langle\sigma_z\rangle$ and the purity $P = \langle\sigma_x\rangle^2 + \langle\sigma_y\rangle^2 + \langle\sigma_z\rangle^2$ from the same dataset, respectively. 

\topic{RB analysis}.
The total error per Clifford $\epsilon$ was extracted by fitting the excited state population $p_e$ by $\langle p_e \rangle = A + B u^m$ where $\langle \cdot \rangle$ denotes the average over a collection of random sequences of $m$ Cliffords, $u = 1 - 2\epsilon$, and $A$ and $B$ are constants determined by state preparation and measurement (SPAM) errors.
The error per gate was then calculated by $\epsilon_g = \epsilon / N$, where $N = 53/24 \approx 2.2083$ was the number of gates per Clifford given our native gateset ${\mathcal{G} = \{I, \pm X_{\pi/2}, \pm Y_{\pi/2}\}}$.

\topic{Purity RB analysis.}
The incoherent error per Clifford $\epsilon_\text{in}$ was extracted by fitting the average state purity $\langle P \rangle = A' + B' u'^m$, where $P = \langle\sigma_z\rangle^2 + \langle\sigma_z\rangle^2 + \langle\sigma_z\rangle^2$ is the state purity extracted by tomography after the $m$-Clifford RB sequence, $u' = (1-2\epsilon_\text{in})^2$, and similarly $A'$ and $B'$ are determined by SPAM errors.
The incoherent error per gate was given by $\epsilon_{g, \text{in}} = \epsilon_\text{in} / N$. 

\begin{figure}[!tb]
\includegraphics[width=0.45\textwidth]{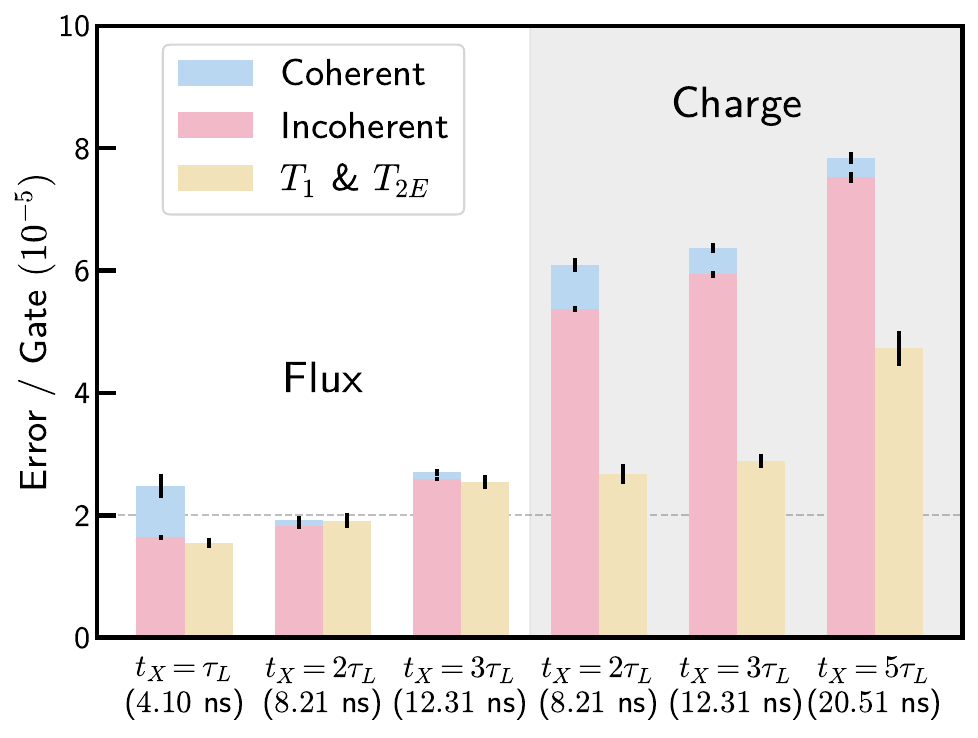}
\caption{
\label{fig:sup_budget}
\textbf{Error budget for commensurate charge and flux gates.}
The total and incoherent error were measured through RB and purity RB, extracted from the same dataset for each gate as described in \aref{app:PRB}.
The estimated incoherent error from decoherence was also extracted from $T_1$ and $T_{2E}$ measurements taken immediately after each RB dataset.
The coherent error budget item was determined by the difference between the total and incoherent errors.
Black lines denote the uncertainty from parameter fits, with the error bar at the top of the coherent budget item corresponding to the total error uncertainty (extracted from the RB fit).
Flux gates (white background) for times $t_X \gtrsim \SI{8}{ns}$ were limited by qubit coherence, whereas charge gates (grey background) were largely limited by incoherent errors beyond undriven qubit decoherence.
}
\end{figure}

We constructed an error budget for several commensurate charge and flux gates by measuring their total and incoherent errors as detailed above and quantifying the proportion of incoherent error from $T_1$ and $T_{2E}$ [\fref{fig:sup_budget}].
Flux gates at times $t_X \gtrsim \SI{8}{ns}$ were limited by incoherent errors, which could be attributed entirely (within uncertainty) to qubit decoherence.
We note that for this dataset, the flux gate set with $t_X = 2\tau_L$ represents our highest measured fidelity with a value of $99.99807(7)\%$.
For our fastest flux gate ($t_X = \tau_L \approx \SI{4}{ns}$), coherent errors were found to play a more significant role, suggesting the breakdown of our calibration.
Charge gates at all times were limited by incoherent errors, and the value could not be entirely accounted for by undriven qubit decoherence. 

\begin{figure}[!tb]
\includegraphics[width=0.45\textwidth]{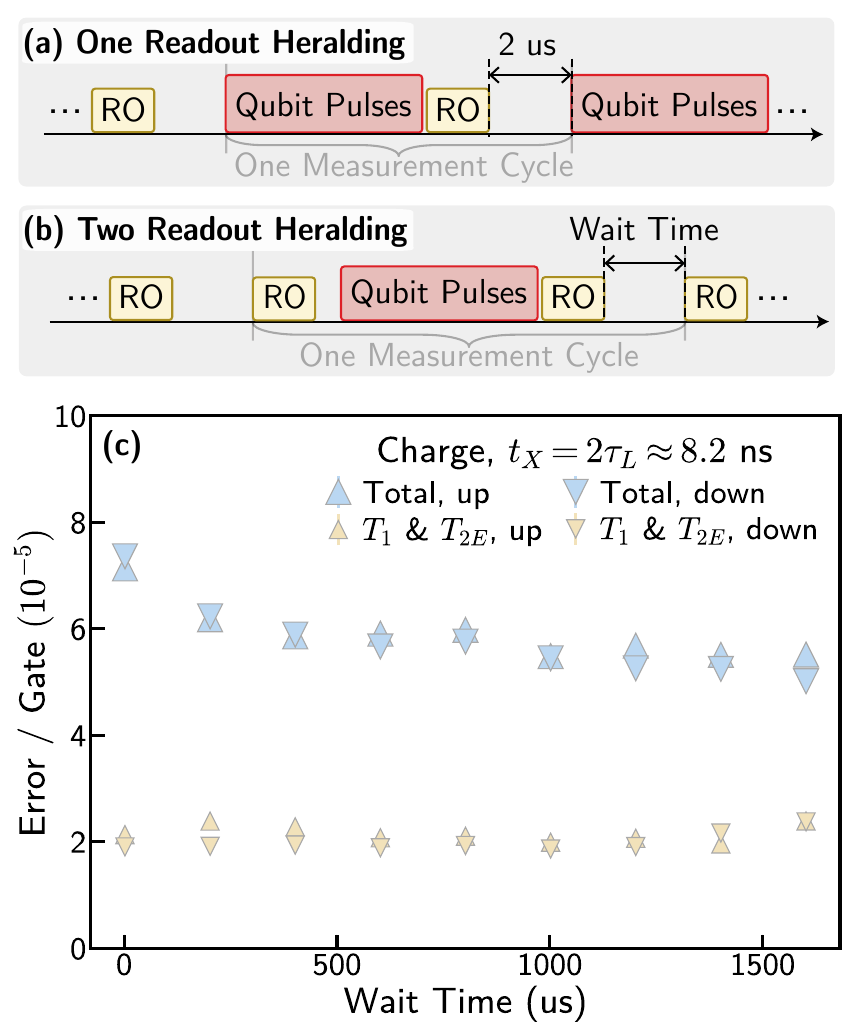}
\caption{
\label{fig:sup_rate}
\textbf{Charge gate heating characterization.}
\cpanel{a} One-readout-pulse heralding scheme, as used for all other data in this work.
\cpanel{b} Two-readout-pulse heralding scheme, as used for the experiment in this figure.
The time between qubit pulses and the preceding readout pulse was kept \SI{2}{\micro s}.
The time between readout pulses was swept in order to investigate the impact of heating on gate performance.
\cpanel{c} Total and $T_1 + T_{2E}$ error for a commensurate charge gate with $t_X = 2\tau_L \approx \SI{8.2}{ns}$ as a function of the wait time between readout pulses, using the pulse sequence of \panel{b}.
The total error (blue) was measured through RB, and the $T_1 + T_{2E}$ error (yellow) was extracted from $T_1$ and $T_{2E}$ measurements taken immediately after each RB dataset.
Data is displayed from back-to-back sweeps of the wait time from 0 to \SI{1600}{\micro s} (upwards triangles) and then \SI{1600}{\micro s} to 0 (downwards triangles).
The total error improved by $20-30\%$ for wait times above \SI{500}{\micro s}, while the measured qubit coherence remained approximately constant.
}
\end{figure}

To probe the role of heating (from, e.g., power dissipated on the chip or in-line attenuators immediately before the device) on charge gate performance, we utilized a two-readout-pulse heralding scheme and swept the wait time between the readout pulses [\fref{fig:sup_rate}].
At long wait times, we expect the relevant components to be in thermal equilibrium with the environment at the start of every qubit pulse sequence and provide no excess decoherence to the qubit.
At short wait times, we expect the elements to heat during an experiment and provide a more severe source of qubit decoherence, which would contribute to incoherent error.
The maximum wait time of \SI{1600}{\micro s} represents a duration of over 5x the longest pulse sequence used in the RB measurements.
The $t_X = 2\tau_L$ commensurate charge gate error displays an improvement until $\approx \SI{500}{\micro s}$ and then plateaus, with a maximum error reduction of $\sim20-30\%$.
The measured coherence times remained approximately constant over the duration of the experiment, supporting our interpretation that the total error trend originated from mitigating system heating.
At the longest wait time (least heating), we still find that our measured decoherence only accounts for at most $\approx 45\%$ of the total error.
We hypothesize that the remaining incoherent error for charge gates is due to leakage, enabled by the large charge matrix elements to higher levels (relative to that of the $\ket{0} \leftrightarrow \ket{1}$ transition). 

\subsection{Stability}
\begin{figure}[!tb]
\includegraphics[width=0.45\textwidth]{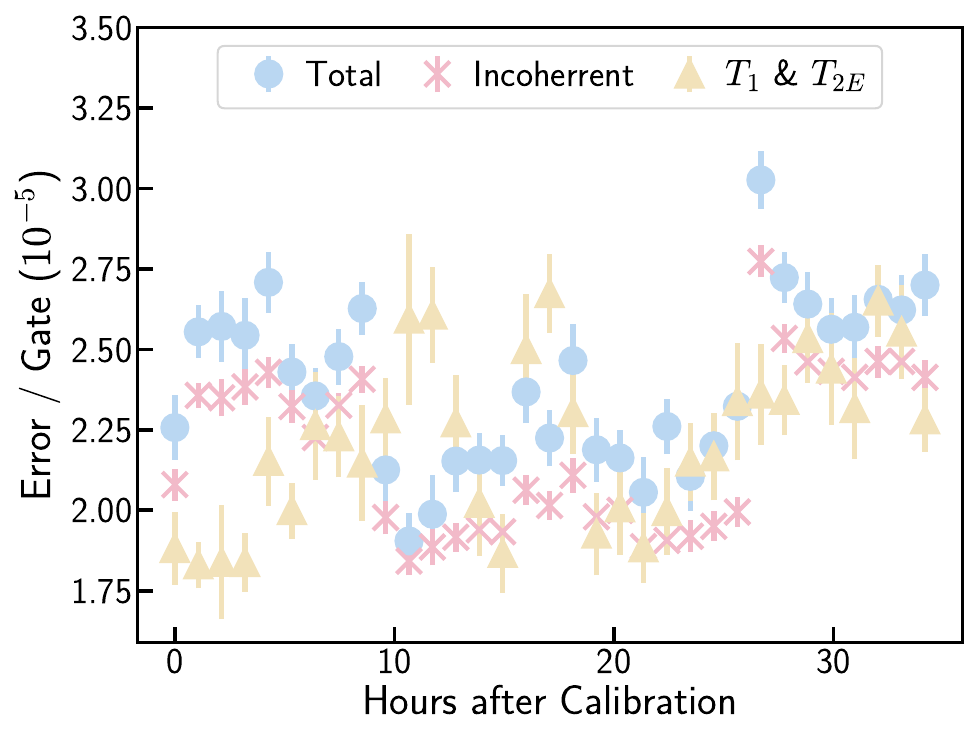}
\caption{
\label{fig:sup_stability}
\textbf{Stability for commensurate flux gates ($t_X = 2\tau_L \approx \SI{8.2}{ns}$).}
The total and incoherent error per gate were measured through RB and purity RB, extracted from the same dataset as described in \aref{app:PRB}.
No significant drift or degredation of gate performance was observed.
}
\end{figure}

To investigate the stability of our system and gate parameters, we measured the total and incoherent error for a commensurate flux gate set with $t_X = 2\tau_L \approx \SI{8.2}{ns}$ (with interleaved $T_1$ and $T_{2E}$ measurements) as a function of time after an initial calibration [\fref{fig:sup_stability}].
Over a span of \SI{34}{hours} after the initial calibration, the total error per gate fluctuated by less than $1.13 \times 10^{-5}$, and displayed no significant degradation.

\section{Comparison of Incommensurate Gates with Linear and Circular Drives}\label{app:incomm_flux_vs_circ}
\begin{figure}[!tb]
\includegraphics[width=0.45\textwidth]{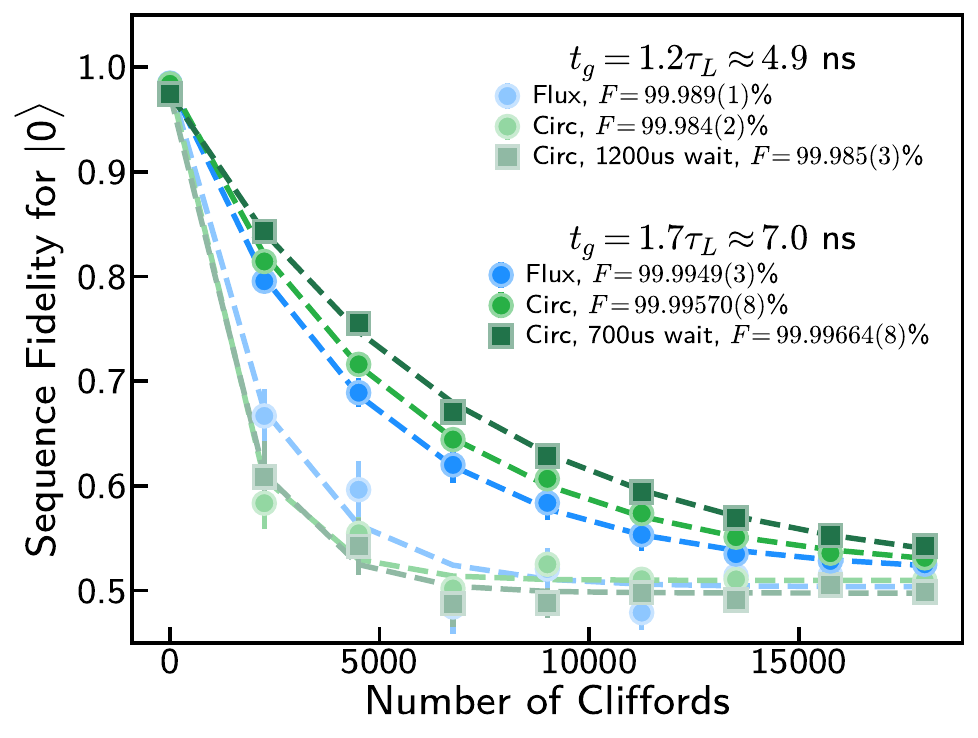}
\caption{
\label{fig:sup_incomm_flux_vs_circ}
\textbf{Comparison of incommensurate gates with linear and circular drives.}
RB data for incommensurate gates utilizing a linearly polarized flux drive and a circularly polarized co-rotating drive is shown for two gate durations, $t_g = 1.2\tau_L \approx \SI{4.9}{ns}$ and $t_g = 1.7\tau_L \approx \SI{7.0}{ns}$.
We include data for circular drives with (square marker) and without (circular marker) additional wait times as defined in \fref{fig:sup_rate} in order to elucidate the efficacy of circular driving for suppressing counter-rotating errors while mitigating heating effects associated with the charge drive.
For $t_g = 1.2\tau_L$, we observed no benefit from circular driving. 
For $t_g = 1.7\tau_L$, circularly polarized driving both with and without an additional wait time benefited gate performance relative to flux driving, which showed no benefit from an equivalent wait time.
}
\end{figure}

Here we present randomized benchmarking data comparing linear and circular driving schemes for incommensurate gates [\fref{fig:sup_incomm_flux_vs_circ}]. 
To represent linear drives, we utilized the best linear driving scheme comprised of a purely flux drive. We present data for two separate gate durations, $t_g = 1.2\tau_L \approx \SI{4.9}{ns}$ and $t_g = 1.7\tau_L \approx \SI{7.0}{ns}$. For $t_g = 1.2\tau_L$, we observe no benefit from circular driving over flux driving, suggesting the relevance of error channels beyond counter-rotating effects as $t_g \rightarrow \tau_L$, e.g., coherent leakage or the non-uniformity of AC Stark shifts. 
In contrast, for $t_g = 1.7\tau_L$, we observe that circular driving outperforms flux driving, signaling the benefit of circular driving for incommensurate single-qubit gates when counter-rotating errors dominate. 

% \subsection{Leakage}

% Leakage outside computational states during a single-qubit gate can be an important error channel when the bandwidth of the control pulse exceeds the qubit anharmonicity. To see if leakage played a major role in limiting our best gate (a flux pulse with duration 10 ns) fidelity, we measured the population of $\ket{2}$ after applying a pulse train $(X_\pi)^N$. 

% The single-shot readout voltages for $\ket{1}$ and $\ket{2}$ overlapped in the IQ plane, making them indistinguishable during regular readout. To overcome this, we took data in two ways: 1) directly after the pulse train, and 2) with an additional $X_\pi$ pulse applied directly before readout. In the first case, the single-shot voltages were classified into two populations $a$ and $b$: $p_a = p_0$ and $p_b = p_1 + p_2$. In the second case, after applying a $\pi$ pulse to the $\ket{0} \rightarrow \ket{1}$ transition, the populations yielded $p_a = p_1$ and $p_b = p_0 + p_2$. We then used the Moore-Penrose pseudoinverse to solve for the populations $p_0$, $p_1$, and $p_2$.

% \subsection{Coherent Errors}

% \section{Error Modelling}

% \subsection{Relaxation into Higher Levels}
% We specifically consider an incoherent relaxation process from $\ket{1}$ to $\ket{2}$ with rate $1/T_{12}$, and derive a gate fidelity limit from this error channel below. An analogous derivation can be made for relaxation from $\ket{0}$ to $\ket{3}$. 

\clearpage

\bibliography{circ_paper}

\end{document}